**Why a Non-Discriminatory Royalty Surcharge Is Not Chip-Neutral:**

**The Error in *FTC v. Qualcomm****

Sang-Seung Yi†

August 30, 2026

**Abstract**

Qualcomm's "No-License, No-Chips" policy let it levy a royalty surcharge on every handset, whether or not it used a Qualcomm modem chip. In *FTC v. Qualcomm*, the Ninth Circuit reversed the district court after accepting Qualcomm's argument that, because the surcharge did not vary with the chip, it was "chip-neutral" and left handset makers' choices undistorted. I develop an equilibrium model of the modem-chip market and show the defense to be wrong: the surcharge's *facial* neutrality does not imply *economic* neutrality. For per-handset surcharges, a surcharge and an equal government tax affect the rival's pricing identically, but not Qualcomm's: a tax is remitted to the Treasury, whereas Qualcomm collects the surcharge — including on handsets using a rival's chip. Raising its own price therefore yields Qualcomm a smaller gain under the surcharge (the surcharge it collects on the demand diverted to the rival) than under the tax (the tax it avoids on its own lost sales), because the diversion ratio is less than one. Under the very conditions that would make a tax chip-neutral, the surcharge raises the rival's all-in price by strictly more than Qualcomm's — and, under symmetric demand, lowers its output by more as well — tilting handset makers toward Qualcomm. For ad-valorem surcharges, the defense fails for a different reason: even a non-discriminatory tax is generically chip-neutral only if the FRAND royalty rate is zero, so the argument's premise itself does not hold. I also analyze discriminatory surcharges.

## 1. Introduction

The Federal Trade Commission's antitrust complaint in *FTC v. Qualcomm*[1] challenged three sets of practices, but at its core was a theory about Qualcomm's "No-License, No-Chips" (NLNC) policy. Under NLNC, Qualcomm refused to sell its modem chips to handset makers that declined its preferred licensing terms, including the royalties it sought on its Standard-Essential Patents (SEPs).[2] The FTC argued that NLNC allowed

* I served as an expert witness adverse to Qualcomm in the Korea Fair Trade Commission's investigations of Qualcomm's licensing business model and in a securities class action against Qualcomm in the United States.

† Department of Economics, Seoul National University, Seoul 08826, Korea. Email: ssyi@snu.ac.kr.

[1] *FTC v. Qualcomm Inc.*, 411 F. Supp. 3d 658 (N.D. Cal. 2019), *rev'd and vacated*, 969 F.3d 974 (9th Cir. 2020).

[2] Complaint at 2, *Qualcomm*, No. 17-cv-00220-LHK (N.D. Cal. Jan. 17, 2017), ECF No. 1 (less-redacted version at ECF No. 38). Qualcomm has committed to standard-setting organizations to license standard-essential patents to all applicants on fair, reasonable, and non-discriminatory ("FRAND") terms. *Qualcomm*, 411 F. Supp. 3d at 671–72.

Qualcomm to leverage its monopoly in modem chips into a surcharge, above the FRAND royalty, on handsets that used a rival's chip — a "naked tax" on transactions between rival chip makers and handset makers that raised rivals' costs and final prices without generating any offsetting efficiency.[3] The district court agreed.[4]

The Ninth Circuit reversed.[5] The tax analogy did not persuade the appellate panel; Qualcomm's "chip-neutrality" defense did. Because Qualcomm imposed the surcharge on *all* handsets — those containing its own chips as well as those containing rivals' — it argued that handset makers' choice between chips was undistorted. The court accepted that argument.[6] The defense equates two distinct notions of neutrality: *facial* neutrality, that the royalty does not discriminate by chip source, and *economic* neutrality, that it leaves chip choice undisturbed.

This paper presents a formal model showing why that argument is wrong. The reasoning is clearest for a per-unit royalty surcharge.[7] The key is to compare a "non-discriminatory" government tax $t$ and a "non-discriminatory" royalty surcharge $s$ of equal magnitude. From the standpoint of the rival chip maker — for example, Intel in the litigation — the two are identical: each reduces Intel's margin by the same amount, and Intel's pricing problem is unchanged.

The difference arises on Qualcomm's side. Let $p_i$ denote firm $i$'s "all-in" price (i.e., surcharge-inclusive) and $q^i(p_1, p_2)$ its demand function, with subscripts on $q^i$ denoting partial derivatives; Qualcomm is firm 1, Intel firm 2. Under a government tax, Qualcomm remits $tq^1$ to the Treasury. Under NLNC, Qualcomm instead collects the surcharge on every handset — including its own, though that component is internalized in its all-in price; what shifts its pricing incentive is the $sq^2$ it collects on the handsets that carry Intel's chips. Even when $s = t$, this changes Qualcomm's marginal incentive to raise its own price. Under the tax, raising $p_1$ by one dollar lets Qualcomm avoid taxes on the $-q_1^1$ units it no longer sells, a saving of $-tq_1^1$. Under the surcharge, raising $p_1$ by one dollar diverts $q_1^2$ units of demand to

[3] *Id*. at 3, 19–21. The FTC also challenged Qualcomm's refusal to license its standard-essential patents to rival chip makers and its exclusivity arrangements with Apple. The refusal to license is a precondition of NLNC: had Qualcomm licensed its SEPs to rival chip makers, the patent-exhaustion doctrine would have foreclosed Qualcomm from demanding a separate royalty from handset makers using those rivals' chips, removing the chip-monopoly leverage that NLNC exploits (*id*. at 22–24). The Apple exclusivity payments include a discriminatory-royalty component of the kind analyzed in Section 7, but they also foreclose rivals from competing for Apple's business — a different mechanism of harm from the cost-raising modeled there; the FTC pleaded them as a distinct claim because of Apple's size and the magnitude of the exclusivity payments (*id*. at 24–28).

[4] *Qualcomm,* 411 F. Supp. 3d at 790–92.

[5] *Qualcomm*, 969 F.3d 974. The district court held that Qualcomm's refusal to license and the Apple exclusivity payments were separate violations of the Sherman Act, *Qualcomm*, 411 F. Supp. 3d at 758–62, 771–72, but the Ninth Circuit reversed those rulings as well, *Qualcomm*, 969 F.3d at 993–95, 1003–05.

[6] *Qualcomm,* 969 F.3d at 1002–03 ("'[N]o license, no chips' is chip-neutral: it makes no difference whether an OEM buys Qualcomm's chip or a rival's chips.").

[7] In *FTC v. Qualcomm*, Qualcomm's royalties had both per-handset and ad-valorem elements. *Qualcomm*, 411 F. Supp. 3d at 673, 715. The per-handset form is the natural starting point and is analyzed in Sections 3–7 (Section 3 sets out the schedule in detail); the ad-valorem case is taken up in Section 8.

Intel, on which Qualcomm now collects an additional $sq_1^2$. The Partial (Derived) Demand Diversion Condition, that the diversion ratio is less than one ($q_1^1 + q_1^2 < 0$; stated as assumption (A3)), implies $sq_1^2 < -tq_1^1$ when $s = t$: the surcharge income gained is less than the tax savings forgone.

Qualcomm therefore has a *weaker* incentive to raise its all-in price under the surcharge than under a government tax of the same magnitude. This comparison exposes the basic flaw in the chip-neutrality argument: it conflates a tax that the Treasury collects with a surcharge that Qualcomm collects, and the identity of the collector matters. Under Qualcomm's implicit premise that a non-discriminatory government tax is chip-neutral, a non-discriminatory royalty surcharge is not. Intel's all-in price rises by more than Qualcomm's, and handset makers are pushed toward Qualcomm's chips.[8]

I formalize the foregoing insight in an equilibrium analysis. The FTC's economics expert, Carl Shapiro, advanced the tax analogy at trial;[9] Shapiro and Waehrer (2023) develop the argument more formally, showing that the seemingly neutral surcharge raises rivals' costs and harms competition. Their account, however, does not explain how the tax and the surcharge differ in their effects on Qualcomm's pricing incentives, nor does it provide an equilibrium analysis of how the surcharge affects modem chip prices, outputs, and profits. As Qualcomm's expert Aviv Nevo put it at trial, the FTC's expert did not present "a complete model of the competitive process."[10] This paper supplies that model. I return to the precise relationship to Shapiro and Waehrer (2023) in Section 6.

Beyond Shapiro and Waehrer (2023), the economics literature on the competitive effects of NLNC is sparse: I have found no formal theoretical analysis of NLNC, either published or in working-paper form. The adjacent literature on patent (including SEP) licensing examines related but distinct questions. Layne-Farrar, Llobet, and Padilla (2014) compare licensing at the component (modem chip) and end-product (handset) levels. Llobet and Padilla (2016) analyze the welfare implications of ad-valorem and per-unit royalties when the patent holder licenses a downstream producer. Padilla and Wong-Ervin (2017) identify conditions under which a refusal to license at the component level is not anticompetitive. Llobet and Neven (2023) show that, although a patent holder generally prefers downstream licensing, this can dampen the end-product maker's incentives to invest. None of these papers addresses the equilibrium effects of a surcharge imposed through modem-chip monopoly leverage of the kind alleged in *FTC v. Qualcomm*.

---

[8] Neutrality in this paper is primarily a statement about the relative all-in price — the chip-choice margin — rather than about quantities, though quantities are addressed as well. Equal increases in the two all-in prices leave that margin undisturbed but still contract handset demand, so neutrality on its own does not pin down the output levels: the comparison of the two firms' outputs is delicate — even under a tax, Qualcomm's own output effect is ambiguous (note 29) — and the conditional chip share can move in either direction with demand asymmetries. The quantity comparison is nonetheless drawn explicitly: Propositions 8 and 9 show that under some conditions the surcharge reduces Intel's output by more than Qualcomm's.

[9] Trial Transcript at 1124, 1137–46, *Qualcomm*, No. 17-cv-00220-LHK (N.D. Cal. Jan. 15, 2019) (testimony of Carl Shapiro).

[10] Trial Transcript at 1900, *Qualcomm*, No. 17-cv-00220-LHK (N.D. Cal. Jan. 25, 2019) (testimony of Aviv Nevo).

The paper proceeds as follows. Section 2 sets out the model and a set of standard assumptions on demand, including the Partial (Derived) Demand Diversion Condition. Section 3 derives the benchmark in which NLNC is prohibited and Qualcomm can collect only the FRAND royalty. Section 4 analyzes a non-discriminatory government tax (Proposition 1). Section 5 replaces the tax with Qualcomm's non-discriminatory royalty surcharge. The all-in prices of both firms' chips rise; under an additional curvature condition (A5′), they rise by less than the surcharge and Intel's margin and profits fall. Qualcomm's margin rises because the non-discriminatory surcharge applies to handsets containing Qualcomm's own chips too, and Qualcomm internalizes that revenue (Proposition 2). I also identify simple sufficient conditions under which Qualcomm's profits rise when it imposes the surcharge (Propositions 3 and 4).

Section 6 contains the paper's central result. The asymmetry described above — Qualcomm's weaker incentive to raise its all-in price under the surcharge than under an equal tax — is formalized in a sequence of propositions. Proposition 5 shows that both firms' all-in prices rise less under a small surcharge than under an equal tax, but the gap is larger for Qualcomm than for Intel. To compare the surcharge to no surcharge at all — the comparison the Ninth Circuit found dispositive — I introduce Qualcomm's Implicit Assumption: that a small non-discriminatory tax raises both firms' all-in prices by the same amount. This assumption is the unstated basis of the chip-neutrality argument. Under it, a small surcharge raises Intel's all-in price by strictly more than Qualcomm's (Proposition 6) and reduces Intel's output (Proposition 7). When one adds the assumption that an equal price increase reduces the two firms' demand by equal amounts (assumption (A4′)), a small surcharge reduces Intel's output by more than Qualcomm's (Proposition 8). Proposition 9 specializes to linear demand with equal own-price and cross-price effects, which satisfies both Qualcomm's Implicit Assumption and (A4′). Linear demand yields closed-form expressions for equilibrium prices and quantities under a surcharge and a tax of any size. These closed forms show that while a non-discriminatory government tax is chip-neutral, a non-discriminatory royalty surcharge is not: Intel's all-in price rises by strictly more than Qualcomm's, and Intel's output falls by strictly more than Qualcomm's.

Section 7 turns to discriminatory surcharges. As a matter of fact, the district court found that Qualcomm offered "chip incentive funds" tied to its own chips to selected handset makers.[11] Shapiro and Waehrer (2023) note that when Qualcomm can adjust its nominal chip price freely, the discriminatory or non-discriminatory character of the surcharge does not affect the competitive outcomes — what matters is the surcharge on handsets containing rival chips. That observation is correct but leaves open why Qualcomm bothered with the incentive funds in the first place. The answer lies in the dynamics of royalty negotiations. New handset makers enter, and existing licenses do not extend to successor technologies (for the period covered by *FTC v. Qualcomm*, 4G LTE emerged as the new global standard). Denote the FRAND royalty rate by $r^F$. If some handset makers, say, shortly after the new standard's introduction, accept Qualcomm's demand to pay $r^F + s$ as a royalty, Qualcomm can argue in subsequent license negotiations that $r^F + s$ has passed a "market test" as a FRAND rate. I present two simple scenarios in which this logic leads Qualcomm to grant lower effective royalties to a subset of customers — a practice that, paradoxically, facilitates the broader imposition of the surcharge.

[11] *Qualcomm*, 411 F. Supp. 3d at 698–702 (LG Electronics), 705, 709 (Samsung Electronics), 714 (Motorola); *see also id.* at 716–745 (further examples).

Section 8 extends the analysis to ad-valorem royalty surcharges, whose base is the handset maker's revenue rather than the number of handsets (and thus chips) sold. The analysis is more complicated, but under reasonable conditions on the pass-through from chip prices to handset prices, the main anticompetitive results carry over. When the Partial (Derived) Demand Diversion Condition is replaced by the stronger Partial Downstream Revenue Diversion Condition, results paralleling Propositions 5 to 8 obtain. One difference is decisive for the chip-neutrality question: even in the simplest symmetric linear case, a non-discriminatory ad-valorem tax is chip-neutral only if the FRAND royalty rate is zero. Because it was undisputed in *FTC v. Qualcomm* that the FRAND royalty rate was positive, the premise underlying the Ninth Circuit's holding fails here as well.

Section 9 concludes.

## 2. The Model

Consider two firms, Qualcomm (firm 1) and Intel (firm 2), competing in prices $p_1$ and $p_2$. Throughout, "Intel" names firm 2 for concreteness, but firm 2 is a stylized stand-in for the non-Qualcomm modem-chip segment[12]: it holds the demand Qualcomm does not, so a higher Qualcomm price diverts buyers to rival chips collectively rather than to any one rival. Representing that segment as a single competitor is a tractability device, with no implication that the rivals coordinate their prices.[13] Demand functions[14] are $q^i(p_1, p_2)$ and costs are linear, $C^i(q^i) = c_i q^i$, so that firm $i$'s profit function is $\pi^i(p_1, p_2) \equiv (p_i - c_i)q^i(p_1, p_2), i = 1, 2$. I keep the same notation for the profit function when the FRAND royalty, tax, and surcharge are introduced in later sections, so that assumption (A5) below holds under all three regimes. Subscripts on demand and profit functions denote partial derivatives; for example, $q^i_j(p_1, p_2) \equiv \frac{\partial q^i(p_1,p_2)}{\partial p_j}$. Throughout the paper, I assume that both firms make positive sales in equilibrium.[15] I impose four assumptions on the first partial derivatives of demand and some technical conditions on the second derivatives.

[12] For the period covered by *FTC v. Qualcomm*, there were several other modem chip rivals, including Broadcom, MediaTek, and Texas Instruments. *Qualcomm*, 411 F. Supp. 3d at 675.

[13] Treating the segment as a single price-setter is a reduced form, not a behavioral claim: the argument uses only that the non-Qualcomm segment has a well-defined aggregate demand and best response and that its aggregate diversion ratio lies below one, the content of Assumption (A3). Any disaggregated model of heterogeneous rivals enters solely through that aggregate ratio, so (A3) is itself the reduced-form aggregation condition, and the single price attributed to the segment is best read as its aggregate price index rather than one rival's quoted price. Deriving (A3) from handset demand, chip qualification, and downstream substitution in a full multi-rival, multi-OEM model is a worthwhile but separate exercise.

[14] To be precise, the demand for modem chips is derived from the demand for handsets that use them. Under the fixed-proportions assumption that one modem chip is installed per handset, this derived nature adds nothing of consequence for per-handset royalties. It matters, however, for ad-valorem royalties on handset prices — a case I defer to Section 8 because of the additional complexities it introduces.

*Negative Own-Price Effect.* Raising a firm's price reduces its own demand:

$$q_i^i(p_1, p_2) < 0, i = 1, 2. \tag{A1}$$

*Positive Cross-Price Effect.* Some of the lost demand is diverted to the rival:

$$q_i^j(p_1, p_2) > 0, i, j = 1, 2, j \neq i. \tag{A2}$$

*Partial Demand Diversion Condition.* The amount diverted to the rival is smaller than the original loss, so industry output falls when one firm raises its price:

$$q_i^i(p_1, p_2) + q_i^j(p_1, p_2) < 0, i, j = 1, 2, j \neq i. \tag{A3}$$

*Negative Effect of a Uniform Price Increase.* If both firms raise their prices by the same amount, each firm's sales fall:

$$q_i^i(p_1, p_2) + q_j^i(p_1, p_2) < 0, i, j = 1, 2, j \neq i. \tag{A4}$$

*Equal Sales Loss under a Uniform Price Increase.* If both firms raise their prices by the same amount, each firm's sales fall by the same amount:

$$q_1^1(p_1, p_2) + q_2^1(p_1, p_2) = q_1^2(p_1, p_2) + q_2^2(p_1, p_2) < 0 \tag{A4$'$}$$

Assumptions (A1)–(A4) are standard in the differentiated-product Bertrand literature. Assumptions (A3) and (A4) are closely related: they coincide when the two cross-price effects are equal ($q_1^2 = q_2^1$). (A4′) is strictly stronger than (A4): (A4) requires only that each firm's sales fall when both prices rise; (A4′) requires those decreases to be equal across the two firms. When the two cross-price effects are equal, (A4′) is satisfied if and only if the own-price effects are also equal ($q_1^1 = q_2^2$).

*Weak Technical Regularity*. The firms' profit functions satisfy the second-order condition (SOC), strategic complementarity, and stability:

$$\pi_{ii}^i < 0, \pi_{ij}^i > 0, \pi_{ii}^i < -\pi_{ij}^i, \qquad i, j = 1, 2, j \neq i. \tag{A5}$$

*Strong Technical Regularity*. The following conditions on the second derivatives of demand strengthen (A5):

$$q_{ii}^i \leq -2q_{ij}^i \leq 0 \; and \; q_{ii}^j \leq -2q_{ij}^j \leq 0, \qquad i, j = 1, 2, j \neq i. \tag{A5$'$}$$

[15] At the end of Section 5, I examine the case in which the two firms incur fixed costs of production. When Intel's operating profits under NLNC do not cover its fixed cost of production, it exits and produces zero output (Corollary 1).

Assumption (A5) is routinely made in the literature.[16] (A5′) implies (A5), given (A1)–(A4) and positive margins. While the best-response slope is less than one under (A5), it is less than one-half under (A5′).

I maintain (A1)–(A5) throughout; the stronger (A4′) and (A5′), which imply (A4) and (A5), respectively, are invoked only where indicated, for the results on margins and profits (Propositions 1(ii), 2(ii), and 3; (A5′)) and on the quantity non-neutrality of a small surcharge (Proposition 8; (A4′)). All of these conditions — including (A4′) and (A5′) — are satisfied by the quasi-symmetric linear demand of Proposition 9 (Section 6), which nests the fully symmetric linear demand of Proposition 4 (Section 5).

Assumption (A3) is the one substantive restriction the mechanism needs, and it is a standard one. In Sections 3 to 7 the differentiated products are the two modem chips, so (A3) says exactly that the inter-chip diversion ratio is below one: when one chipmaker raises its all-in price, the sales it loses exceed the sales the rival gains, and total chip output falls. Equivalently, the own-price effect dominates the cross-price effect — the standard regularity condition on differentiated-products demand (Singh and Vives 1984, Section 5). With more than one rival, (A3) is the condition that the *aggregate* diversion ratio[17] — the fraction of Qualcomm's lost sales recaptured by all rival chips together — is below one. It is below one because a higher modem price is passed through, in part, to a higher price for the handsets carrying that chip: some buyers switch to rival-chip handsets — the diversion the cross-price term records — while others postpone replacing their handsets or stop buying new ones altogether rather than pay more.

## 3. Benchmark: NLNC prohibited

As a benchmark, suppose NLNC is prohibited, so Qualcomm can collect only the FRAND royalty from handset makers. For the period covered by *FTC v. Qualcomm*, Qualcomm's royalty structure varied with the identity of the handset maker and the price of the handset. Most notably, it charged Apple a fixed \$7.50 per handset.[18] From other handset makers, it typically charged a percentage of the handset price (for example, 5% on 3G handsets, including 3G-4G "multimode" handsets), subject to a cap of \$20.[19] Hence, for handsets with a net selling price above \$400, the royalty was fixed at \$20 regardless of price — effectively a per-handset royalty. For simplicity, in Sections 3 to 7 I examine a per-handset royalty.[20] The observed terms — the Apple charge and the \$20 cap — are cited as evidence of this per-handset form, not as the FRAND benchmark itself: $r^F$ denotes the hypothetical rate Qualcomm could charge absent NLNC, and the surcharge $s$ is the increment of the actual charge above that benchmark.

[16] *See* Whinston (1990), for example.
[17] The aggregate diversion ratio, or the recapture ratio, plays a central role in merger analysis. See Farrell and Shapiro (2010) and DOJ and FTC's Merger Guidelines (2023). For simplicity, in this paper, I use the term diversion ratio.
[18] *Qualcomm*, 411 F. Supp. 3d at 725 ("Apple viewed the \$7.50 per handset royalty payment the companies settled on as excessive, Apple had no alternative.").
[19] *Id.* at 673. Qualcomm also charged an upfront fee in many licensing contracts (in addition to the running royalty), but it does not affect the analysis.
[20] In Section 8, I extend the analysis to ad-valorem royalty/surcharge on handset prices.

Let $r^F$ denote the per-handset FRAND royalty and let $p_i^F$ denote the "all-in" price — the sum of the nominal chip price and the FRAND royalty. The royalty enters the two firms' profit maximization problems asymmetrically, for two reasons. First, Intel's profit margin on its chips is reduced by $r^F$, while Qualcomm's is not. Second, Qualcomm collects $r^F$ for each handset containing an Intel chip.

Qualcomm's profit is

$$\pi^1(p_1^F, p_2^F) = (p_1^F - c_1)q^1(p_1^F, p_2^F) + r^F q^2(p_1^F, p_2^F) \qquad (Eq\ 3.1)$$

while Intel's profit is

$$\pi^2(p_1^F, p_2^F) = (p_2^F - c_2)q^2(p_1^F, p_2^F) - r^F q^2(p_1^F, p_2^F) \qquad (Eq\ 3.2)$$

The equilibrium all-in prices (inclusive of FRAND royalty), denoted by $(p_1^{F*}, p_2^{F*})$, satisfy the two firms' first-order conditions simultaneously:

$$\pi_1^1(p_1^F, p_2^F) = q^1(p_1^F, p_2^F) + (p_1^F - c_1)q_1^1(p_1^F, p_2^F) + r^F q_1^2(p_1^F, p_2^F) = 0 \qquad (Eq\ 3.3)$$

and

$$\pi_2^2(p_1^F, p_2^F) = q^2(p_1^F, p_2^F) + (p_2^F - c_2 - r^F)q_2^2(p_1^F, p_2^F) = 0 \qquad (Eq\ 3.4)$$

Under (A1)–(A2) and positive equilibrium sales, both firms have market power: equilibrium all-in prices exceed marginal costs (in Intel's case, marginal cost plus the FRAND royalty), and each earns positive economic profits. Rearranging the FOCs yields

$$p_1^{F*} - c_1 = -\frac{q^1(p_1^{F*}, p_2^{F*}) + r^F q_1^2(p_1^{F*}, p_2^{F*})}{q_1^1(p_1^{F*}, p_2^{F*})} > 0 \qquad (Eq\ 3.5)$$

$$p_2^{F*} - c_2 - r^F = -\frac{q^2(p_1^{F*}, p_2^{F*})}{q_2^2(p_1^{F*}, p_2^{F*})} > 0 \qquad (Eq\ 3.6)$$

Because Intel earns positive economic profits in this benchmark, Qualcomm may have an incentive to weaken Intel — for instance, by raising its effective marginal cost via NLNC.[21] Doing so loosens the competitive constraint that Intel imposes on Qualcomm's pricing, and, over a longer horizon, on the two firms' investment decisions. Whether weakening Intel is on balance profitable is not automatic, however — Qualcomm also earns

[21] For a general discussion of raising-rivals'-costs strategy employed by dominant firms, *see* Krattenmaker and Salop (1986).

FRAND royalties on Intel-chip handsets — and the later propositions establish the conditions under which it is.

## 4. Non-Discriminatory Government Taxation

Before turning to Qualcomm's surcharge, consider a counterfactual: a per-unit tax $t$ levied by the government on every handset, regardless of which chip it contains. Let $p_i^G$ denote firm $i$'s tax-inclusive all-in price. The two firms' profits are

$$\pi^1(p_1^G, p_2^G) = (p_1^G - c_1)q^1(p_1^G, p_2^G) + r^F q^2(p_1^G, p_2^G) - tq^1(p_1^G, p_2^G) \quad (Eq\ 4.1)$$

$$\pi^2(p_1^G, p_2^G) = (p_2^G - c_2)q^2(p_1^G, p_2^G) - (r^F + t)q^2(p_1^G, p_2^G) \quad (Eq\ 4.2)$$

The equilibrium all-in prices (inclusive of FRAND royalty and tax), denoted by $(p_1^{G*}, p_2^{G*})$, satisfy the two firms' first-order conditions simultaneously:

$$\pi_1^1(p_1^G, p_2^G) = q^1 + (p_1^G - c_1)q_1^1 + r^F q_1^2 - tq_1^1 = 0 \quad (Eq\ 4.3)$$

and

$$\pi_2^2(p_1^G, p_2^G) = q^2 + (p_2^G - c_2)q_2^2 - (r^F + t)q_2^2 = 0 \quad (Eq\ 4.4)$$

Proposition 1 records the standard incidence and pass-through results for a uniform tax in this differentiated-product Bertrand setting.

**Proposition 1.** *(i) Assume (A1)–(A5). A non-discriminatory per-handset tax raises both firms' equilibrium all-in modem chip prices and reduces total output.*[22]

*(ii) If in addition (A5′) holds, both prices rise by less than the tax, so both firms' margins and profits fall.*

All proofs appear in the Appendix (except Proposition 9, where I use the closed-form solutions for equilibrium prices under the tax and surcharge to draw Figure 1). Proposition 1 records what is already familiar from the public-finance literature: a non-discriminatory tax raises prices and reduces total output under standard assumptions. When additional mild assumptions on the curvature of the demand functions in the form of (A5′) are added, an

[22] Although the proof signs the effect of an infinitesimal tax, Propositions 1 and 2 are global comparisons of any tax with none: the marginal effect keeps its sign along the path from the no-tax equilibrium, so integrating it yields the cumulative effect. This presumes the equilibrium remains interior, with both firms selling positive quantities.

*industry-wide* cost shock (such as a per-unit tax) passes through only partially, so both margins and profits fall. This section establishes the benchmark against which Qualcomm's surcharge — formally similar but institutionally different — is compared in Section 5.

A remark on the output effect of an industry-wide tax is in order. At this level of generality, one cannot determine whose price rises more. As shown in the proof of Proposition 1, the firm whose price rises more sells fewer modem chips. But the complexities of oligopolistic competition leave open the possibility that the other firm sells more. The same caveat applies to a non-discriminatory royalty surcharge, but I identify conditions under which Intel's output falls.

## 5. No-License, No-Chips

Replace the government tax with a per-handset surcharge $s$ imposed by Qualcomm itself, on top of the FRAND royalty $r^F$ and applied uniformly across handsets — those with Qualcomm chips and those with rivals'. The surcharge is the policy instrument NLNC enables: by tying chip supply to acceptance of its preferred royalty terms, Qualcomm forces handset makers to pay $r^F + s$ regardless of whose chips they use.

Let $p_i^N$ denote firm $i$'s surcharge-inclusive all-in price; the superscript $N$ stands for NLNC. The two firms' profits are

$$\pi^1(p_1^N, p_2^N) = (p_1^N - c_1)q^1(p_1^N, p_2^N) + (r^F + s)q^2(p_1^N, p_2^N) \qquad (Eq\ 5.1)$$

$$\pi^2(p_1^N, p_2^N) = (p_2^N - c_2)q^2(p_1^N, p_2^N) - (r^F + s)q^2(p_1^N, p_2^N) \qquad (Eq\ 5.2)$$

The equilibrium all-in prices (inclusive of FRAND royalty and surcharge), denoted by $(p_1^{N*}, p_2^{N*})$, satisfy the two firms' first-order conditions simultaneously:

$$\pi_1^1(p_1^N, p_2^N) = q^1(p_1^N, p_2^N) + (p_1^N - c_1)q_1^1(p_1^N, p_2^N) + (r^F + s)q_1^2(p_1^N, p_2^N) = 0 \quad (Eq\ 5.3)$$

$$\pi_2^2(p_1^N, p_2^N) = q^2(p_1^N, p_2^N) + (p_2^N - c_2)q_2^2(p_1^N, p_2^N) - (r^F + s)q_2^2(p_1^N, p_2^N) = 0 \quad (Eq\ 5.4)$$

The structural difference between the tax and the surcharge is institutional, but it has a sharp algebraic counterpart in Qualcomm's problem: the marginal policy term in its first-order condition shifts from $-tq_1^1$ (Eq 4.3) to $+sq_1^2$ (Eq 5.3). Instead of remitting tax revenue to the Treasury, Qualcomm now collects the surcharge from Intel as well as from itself. From Intel's standpoint, however, the surcharge looks exactly like a tax. Under the additional curvature condition (A5′) (Proposition 2(ii)), Intel's all-in price rises by less than the surcharge, its margin $p_2^{N*} - c_2 - r^F - s$ falls, and its profits decline.

Qualcomm's situation is different. Like a tax, a surcharge exerts upward pricing pressure (UPP) on Qualcomm's price, but for a different reason: when Qualcomm raises its price, some of its sales are diverted to Intel, on which Qualcomm collects the surcharge. As a result, Qualcomm raises its all-in price; but because Qualcomm itself collects the surcharge, its margin $p_1^{N*} - c_1$ rises. Both firms' all-in prices increase, so handset makers — and ultimately final consumers — pay more.[23] The conduct is anticompetitive: Qualcomm exploits its chip monopoly to impose a surcharge on transactions between rivals and handset makers.

**Proposition 2.** *(i) Assume (A1)–(A5). A non-discriminatory per-handset royalty surcharge raises both firms' equilibrium all-in modem chip prices and reduces total output; Qualcomm's margin rises, because it collects the surcharge on its own chips as well.*

*(ii) If in addition (A5′) holds, both prices rise by less than the surcharge, so Intel's margin and profits fall.*

The analysis so far has abstracted from fixed costs. For modem chips, the fixed costs of R&D, fabrication, and certification are substantial.[24] Let $K_i$ denote firm $i$'s fixed cost, and let $\pi^i(p_1^{F*}, p_2^{F*})$ and $\pi^i(p_1^{N*}, p_2^{N*})$ denote firm $i$'s operating profits in the FRAND and NLNC equilibria respectively. Proposition 2(ii) establishes that, under the additional curvature condition (A5′), $\pi^2(p_1^{N*}, p_2^{N*}) < \pi^2(p_1^{F*}, p_2^{F*})$: a sufficiently large surcharge can therefore drive Intel below break-even.

**Corollary 1.** *If $\pi^2(p_1^{N*}, p_2^{N*}) < K_2 < \pi^2(p_1^{F*}, p_2^{F*})$, NLNC induces Intel's exit, leaving Qualcomm to charge the monopoly price.*[25]

Proposition 2 is silent on Qualcomm's profits, for which the analysis is more delicate: a higher surcharge raises Qualcomm's per-unit revenue from rival handsets but also reduces

[23] Formally, an increase in the royalty surcharge has the same effect on equilibrium prices as an equal increase in the FRAND royalty, because only the total royalty matters for the pricing equations. See the first-order conditions (Eq 5.3) and (Eq 5.4). But as Shapiro and Waehrer (2023, footnote 11) observe, the FRAND royalties "by definition reflect the reasonable value of Qualcomm's SEP portfolio, which itself reflects Qualcomm's R&D investments that led to those patents." What is challenged is Qualcomm's exercise of monopoly power in modem chips to impose a royalty surcharge on top of FRAND royalties on handsets with rival chips.

[24] *Qualcomm*, 411 F. Supp. 3d at 688–89, 695.

[25] A simple example illustrating the corollary's logic, even outside the model's assumptions, is homogeneous-product Bertrand competition with demand function $Q(p)$. (Note that (A3), the diversion-ratio assumption, fails when the two firms charge identical prices, since a small price increase by one firm shifts all of its demand to the rival.) Assume neither firm charges below its marginal cost (including the royalty for Intel). If $c_2 + r^F < c_1 < c_2 + r^F + s$, then absent the surcharge Intel wins by charging $c_1 - \varepsilon$, while Qualcomm prices at $c_1$. The surcharge inverts this: Qualcomm wins by charging $c_2 + r^F + s - \varepsilon$. If $(c_1 - r^F - c_2)Q(c_1) > K_2$, Intel earns a positive net profit absent the surcharge but exits when the surcharge is imposed. After exit, Qualcomm prices at the unconstrained monopoly level $p_1^M$ that maximizes $(p - c_1)Q(p)$.

$q^2$, the base on which it collects the surcharge (as well as the FRAND royalty), so the net effect depends on the magnitudes. The next two propositions identify conditions under which the surcharge raises Qualcomm's profits.

Proposition 3 gives a local result: under (A1)–(A4) and (A5′), the surcharge raises Qualcomm's profit at the margin whenever the total royalty $r^F + s$ is small relative to Intel's gross margin. Proposition 4 gives a global result for a symmetric linear demand: imposing any surcharge $s$ is more profitable than no surcharge, provided $r^F$ is not too large. The common theme is that when the FRAND royalty is small, Qualcomm benefits from imposing a surcharge under NLNC.

**Proposition 3.** *Assume (A1)–(A4) and (A5′). Qualcomm's marginal profit is positive if the total per-handset royalty is less than half Intel's gross margin:*

$$\frac{d\pi^1(p_1^{N*}, p_2^{N*})}{ds} > 0 \; if \; r^F + s < \frac{p_2^{N*} - c_2}{2}.$$

As the diversion ratio approaches one, the exact condition (Eq 5.5) — which retains the pass-through term $\frac{dp_2^{N*}}{ds}$ — is approximated by the simpler bound

$$r^F + s < max\left\{p_1^{N*} - c_1, \frac{(p_1^{N*} - c_1) + (p_2^{N*} - c_2)}{2}\right\}$$

Proposition 3 is a *local* result: it identifies when the surcharge raises profits at the margin. Even when Proposition 3's conditions fail — so that further increases would reduce profits — imposing some surcharge can still be more profitable than imposing none. To examine this *global* comparison, the next proposition specializes to a symmetric linear demand (and symmetric cost).

**Proposition 4.** *Consider the symmetric linear demand for differentiated products* [26]

$$q^i(p_1, p_2) = \frac{1}{1 - d^2}\left[(1 - d)A - p_i + dp_j\right], 0 < d < 1, i = 1, 2$$

*with $c_1 = c_2 = c$ and $A > c$. Imposing a non-discriminatory royalty surcharge $s$ raises Qualcomm's profit relative to no surcharge if and only if*[27]

[26] The symmetric linear demand of Proposition 4 is the special case of the quasi-symmetric demand in Proposition 9 with equal intercepts, $a_i = (1 - d)A$; see footnote 36 for the generating utility.

[27] For Proposition 4's comparison to be relevant, the equilibrium must be interior. The Appendix shows that both firms sell positive quantities in equilibrium if and only if $r^F + s < \frac{(2+d)}{2(1+d)}(A - c)$. It is easy to confirm that $\frac{(2+d)}{2(1+d)} < \frac{8+d^3}{8+d^2}$. Hence, if $r^F \approx 0$, Qualcomm earns a higher profit under any surcharge than under no surcharge in any interior equilibrium.

$$2r^F + s < \frac{8 + d^3}{8 + d^2}(A - c).$$

### 6. The Ninth Circuit's Error

Proposition 2 establishes that, from Intel's standpoint, Qualcomm's surcharge acts like a government tax: Intel's effective marginal cost rises, and its all-in price rises; under the additional curvature condition (A5′), its margins and profits both fall (Proposition 2(ii)), while the chip-neutrality comparison that follows uses only (A1)–(A5).

Intel's weakened position lets Qualcomm raise its own all-in price and earn higher margins. Both increases pass through, in part, to final consumers as higher handset prices — without any offsetting public revenue.

The FTC and its economics expert advanced this argument at trial, and the district court accepted it.[28] The Ninth Circuit, however, found the tax analogy insufficient. The appellate panel was persuaded instead by Qualcomm's defense that the surcharge was "chip-neutral."[29]

This section provides the formal critique of the chip-neutrality argument. I first show in Proposition 5 that both firms' all-in prices rise less under a small surcharge than under an equal tax, but that the difference between the tax-induced and surcharge-induced price increases is larger for Qualcomm than for Intel. As a result, *compared to a tax*, a surcharge induces handset makers to choose Qualcomm's modem chips over Intel's. I then introduce Qualcomm's Implicit Assumption — the assumption that a (small) non-discriminatory tax is chip-neutral in price — and show that under this assumption, a small surcharge, *compared to no surcharge*, increases Intel's price by more than Qualcomm's (Proposition 6) and reduces Intel's output (Proposition 7). Under (A4′), which strengthens (A4) by requiring equal sales loss across firms, Proposition 8 establishes the comparative quantity effect: if Qualcomm's output falls, Intel's falls by more.[30] Proposition 9 then provides a *global* result for a quasi-symmetric linear demand system.

The key insight rests on a similarity and a difference between a government tax and Qualcomm's surcharge.

*The similarity.* When $s = t$, Intel's pricing problem under Qualcomm's surcharge is identical to that under a government tax. Comparing (Eq 4.4) with (Eq 5.4): both reduce Intel's margin by the same amount, so Intel's FOC is the same under either regime when evaluated at the same prices.

[28] *Qualcomm*, 411 F. Supp. 3d at 790–92.
[29] *Qualcomm*, 969 F.3d at 1002–03 (quoted *supra* note 6).
[30] Assumption (A4′) (Equal Sales Loss) is restrictive in asymmetric markets, and it enters only the comparative-output result (Proposition 8). The paper's central claim — that the surcharge tilts the relative all-in price and so is not chip-neutral — rests on the relative-price results of Sections 5–6, which do not invoke (A4′); relaxing it would qualify the ranking of output losses without disturbing that conclusion.

*The difference.* Under a tax, Qualcomm remits $tq^1$ to the Treasury; under the surcharge, Qualcomm collects $sq^2$ from handset makers on the handsets that carry Intel's chips. This changes Qualcomm's marginal incentive to raise its own price even when $s = t$. Comparing (Eq 4.3) with (Eq 5.3): under the tax, raising $p_1$ by one dollar saves Qualcomm $-tq_1^1$ in avoided tax on units no longer sold; under the surcharge, raising $p_1$ by one dollar yields an additional $sq_1^2$ in royalty income from units diverted to Intel. The Partial Demand Diversion Condition (A3), $q_1^1 + q_1^2 < 0$, implies $sq_1^2 < -tq_1^1$ when $s = t$: the surcharge income gained falls short of the tax savings forgone.

*The upshot*: Qualcomm has a *weaker* incentive to raise its all-in price under the surcharge than under a tax of equal magnitude. This is where the chip-neutrality argument fails. Qualcomm's defense conflates a government tax (which the Treasury collects) with a surcharge that Qualcomm itself collects — and the institutional difference matters. Even if a tax raises both firms' all-in prices by the same amount, the surcharge raises Intel's by more than Qualcomm's, biasing handset makers toward Qualcomm's chips. Proposition 5 makes the surcharge–tax comparison precise.

**Proposition 5.** *Assume (A1)–(A5). Starting from $s = t = 0$, suppose a small surcharge of size $s$ is imposed in one regime and a tax of equal size $t = s$ in the other. Both firms' all-in prices rise less under the surcharge than under the tax, but the gap is larger for Qualcomm than for Intel:*

$$0 < \left.\frac{dp_2^{G*}}{dt}\right|_{t=0} - \left.\frac{dp_2^{N*}}{ds}\right|_{s=0} < \left.\frac{dp_1^{G*}}{dt}\right|_{t=0} - \left.\frac{dp_1^{N*}}{ds}\right|_{s=0}$$

That is, the surcharge's price increase falls short of the tax-induced increase by more for Qualcomm than for Intel.

Proposition 5 compares the surcharge regime to the *tax* regime, but does not directly address the chip-neutrality argument the Ninth Circuit found dispositive: it accepted Qualcomm's claim that the surcharge leaves the choice of chips unaffected. To compare the surcharge regime to the *no-surcharge* regime, I introduce an additional assumption: a small non-discriminatory tax raises both firms' all-in prices by the same amount.[31] Under this assumption, a non-discriminatory tax can be called "chip-neutral", because, to paraphrase Qualcomm, "the [tax] does not push the [handset makers] in either direction."[32] I call this assumption Qualcomm's Implicit Assumption, because it is the unstated basis for the "chip-neutrality" argument.

[31] In general, a non-discriminatory government tax can have differential effects on the equilibrium all-in prices of oligopolistic firms when the slopes of the two firms' best-response functions differ (even when $\frac{q_1^1}{\pi_{11}^1} = \frac{q_2^2}{\pi_{22}^2}$ so that as a hypothetical monopolist, each firm would increase its price by the same amount; *see* footnote 52 in the Appendix). One can see this by comparing $\frac{dp_1^{G*}}{dt}$ and $\frac{dp_2^{G*}}{dt}$ in the proof of Proposition 1.

[32] Reply Brief for Appellant Qualcomm at 29, *FTC v. Qualcomm Inc.*, 969 F.3d 974 (9th Cir. 2020) (No. 19-16122), Dkt. Entry 228 (Dec. 16, 2019).

*Qualcomm's Implicit Assumption*

$$\left.\frac{dp_1^{G*}}{dt}\right|_{t=0} = \left.\frac{dp_2^{G*}}{dt}\right|_{t=0}$$

Under this assumption, a non-discriminatory tax is chip-neutral in the sense that it raises both firms' prices by the same amount. Qualcomm's brief invokes precisely this intuition and applies it to the total royalty (which is the FRAND royalty plus the surcharge):

> "[T]he choice of chip does not change the Qualcomm royalty. Thus, an OEM can make its chip choice based solely on the price and quality of chips; the royalty does not push the OEM in either direction. Instead, all chip manufacturers including Qualcomm compete for the OEM's business on an even footing."[33]

Adding Qualcomm's Implicit Assumption to the other assumptions of Proposition 5 establishes that the surcharge — contrary to Qualcomm's chip-neutrality claim — raises Intel's all-in price by more than Qualcomm's, "pushing the OEM in [Qualcomm's] direction".

**Proposition 6 (Price Non-Neutrality of a Small Surcharge).** *Assume (A1)–(A5) and Qualcomm's Implicit Assumption. Starting from no surcharge, the marginal effect of a non-discriminatory royalty surcharge is to raise Intel's all-in price by more than Qualcomm's:*

$$0 < \left.\frac{dp_1^{N*}}{ds}\right|_{s=0} < \left.\frac{dp_2^{N*}}{ds}\right|_{s=0}$$

Consider quantities, beginning with Intel. Its own all-in price rises, reducing demand for its chips; but Qualcomm's price rises too, offsetting part of that reduction. Best-response interaction in oligopoly means that, for general demand systems, it is difficult to identify simple conditions under which a surcharge reduces Intel's output.[34] Under Qualcomm's Implicit Assumption, however, a clean result obtains.

---

[33] *Id.* at 28–29.

[34] By Intel's first-order condition (Eq 5.4), I have $q^2(p_1^N, p_2^N) = -(p_2^N - c_2 - r^F - s)q_2^2(p_1^N, p_2^N)$. In the case of linear demand, $q_2^2$ is a constant. Hence, Intel's equilibrium output is proportional to its margin $p_2^N - c_2 - r^F - s$. Since the surcharge reduces this margin (Proposition 2) for any $s > 0$, it

**Proposition 7 (Rival's Output Reduction under a Small Surcharge).** *Assume (A1)–(A5) and Qualcomm's Implicit Assumption. Starting from no surcharge, the marginal effect of a non-discriminatory royalty surcharge is to reduce Intel's output:*

$$\left.\frac{dq_2^{N*}}{ds}\right|_{s=0} < 0$$

Turning to Qualcomm, the comparison is more delicate because of the royalty income.[35] Qualcomm's own price increase reduces its output, while Intel's price increase shifts demand toward Qualcomm; the two forces work in opposite directions, so the net effect on Qualcomm's output is generally ambiguous.[36] Under Qualcomm's Implicit Assumption, Intel's price rises more than Qualcomm's, so intuition suggests that if Qualcomm's output falls, Intel's falls by more. The comparison again has hidden subtleties — this time requiring a strengthening of (A4) to (A4′).

**Proposition 8 (Quantity Non-Neutrality of a Small Surcharge).** *Assume (A1)–(A3), (A4′), (A5) and Qualcomm's Implicit Assumption. Starting from no surcharge, the marginal effect of a non-discriminatory royalty surcharge satisfies:*

$$\left.\frac{dq_2^{N*}}{ds}\right|_{s=0} < \left.\frac{dq_1^{N*}}{ds}\right|_{s=0}$$

*In particular, if Qualcomm's output decreases under the surcharge, Intel's output decreases by strictly more.*[37]

---

reduces Intel's output for any $s > 0$ — extending Proposition 7's local result to a global one for linear demand. This argument is more general than Proposition 9's — it applies to *any* linear demand satisfying (A1)–(A5) — but Proposition 9 establishes the additional comparative result that the surcharge reduces Qualcomm's output by less than Intel's under symmetric own-price and cross-price effects of Proposition 9's demand system.

[35] Under NLNC, Qualcomm collects the surcharge (on top of the FRAND royalty) on Intel's chips (to be more precise, handsets containing Intel's chips). The surcharge's effect on Qualcomm's output therefore depends on more than its own margin; see footnote 34 for the formal treatment.

[36] For a linear demand, I could establish that the surcharge reduces Intel's equilibrium output. *See supra note* 32. For Qualcomm, however, even for linear demand it is unclear if the surcharge will reduce its output. To see why, rearrange Qualcomm's first-order condition (Eq 5.3) to obtain $q^1(p_1^N, p_2^N) = -(p_1^N - c_1)q_1^1(p_1^N, p_2^N) - (r^F + s)q_1^2(p_1^N, p_2^N)$. Due to the presence of the per-unit royalty $r^F + s$, Qualcomm's equilibrium output is *not* proportional to its margin $p_1^N - c_1$. Hence, the simple logic in footnote 32 does not apply to Qualcomm.

[37] Due to the complexities of feedback effects through best-response interaction, one cannot rule out that Qualcomm's equilibrium output increases. That is why the qualifier "if Qualcomm's output decreases" is added. In the quasi-symmetric linear demand system considered in Proposition 9, however, Qualcomm's equilibrium output decreases.

Two results should be kept distinct. Proposition 5 is a relative statement that needs no appeal to Qualcomm's Implicit Assumption: compared with an equal tax, the surcharge tilts chip choice toward Qualcomm. On the court's own premise that a non-discriminatory tax is neutral, this alone contradicts its holding that the surcharge leaves chip choice undistorted: a surcharge that tilts choice relative to a neutral tax cannot itself be neutral. Propositions 6 to 8 deliver the stronger, absolute statement, that the surcharge raises Intel's price and lowers its output relative to no surcharge, but only under Qualcomm's Implicit Assumption. The refutation therefore does not stand or fall with that assumption: the relative result holds regardless, and the absolute result holds on Qualcomm's own premise.

Within those absolute results, the dependence on symmetry is uneven. The comparative output result, that Intel's output falls by strictly more than Qualcomm's, is the one conclusion that also needs the symmetry condition (A4′) (or the quasi-symmetric linear demand of Proposition 9); the price non-neutrality of Proposition 6 and the reduction in Intel's output of Proposition 7 do not.

A final word on the relative refutation. Proposition 5 itself takes no stand on whether either charge is neutral; the step from it to the holding's falsity — that a surcharge which tilts choice relative to a tax cannot itself be neutral — does lean on one premise, that the tax it is measured against is neutral, and one might ask what survives in a market asymmetric enough that even a non-discriminatory tax is not chip-neutral. That reliance can be removed. Write the surcharge's effect on the gap between the two firms' all-in prices as the effect an equal tax would have on that gap, plus the increment Proposition 5 isolates. Qualcomm's Implicit Assumption fixes only the first term, setting it to zero; the second is unconditional, favoring Qualcomm whenever the diversion ratio is below one (Assumption (A3)). When the first term is not zero — an asymmetric tax that is not itself neutral — the surcharge does not merely inherit that distortion and stop; it adds the Proposition 5 increment on top, always in Qualcomm's favor. That increment is precisely the surcharge's departure from the tax it is likened to: it falsifies the likeness itself — the step on which the defense rests — with no premise about whether the tax is neutral.

It helps to separate what the model shares with the existing account of the case from what it adds. The anticompetitive character of the NLNC tie is shared ground: conditioning chip supply on a per-handset license raises rivals' costs, weakens Intel and can drive it from the market (Proposition 2 and Corollary 1), and lets Qualcomm raise its own all-in price and earn higher profits (Propositions 3 and 4). This is the tax theory Carl Shapiro advanced at trial and that Shapiro and Waehrer (2023) develop; the credit for it is theirs, and Propositions 2 to 4 do no more than recast it as the equilibrium of an explicit model — the "complete model of the competitive process" the trial testimony was faulted for lacking. The model therefore establishes anticompetitive harm, not merely non-neutrality.

The paper's own contribution lies in Propositions 5 to 9 and their Appendix B counterparts, which meet the defense that prevailed on appeal rather than the FTC's theory of harm. Qualcomm's surcharge is neutral on its face — non-discriminatory, applied to every handset whatever chip it carries — and the chip-neutrality holding inferred from that facial neutrality that the surcharge is neutral in economic effect, leaving the competitive playing field level. The inference fails. A non-discriminatory surcharge is not the non-discriminatory tax it resembles: it bears on the rival's price as a tax would but returns to Qualcomm only the royalty on the sales it diverts to the rival, so Qualcomm's incentive to raise its own all-in price is the weaker of the two, the rival's price rises by more, and chip choice is distorted (Propositions 5 to 8, with Proposition 9 supplying the global comparison). *Facial* neutrality does not imply *economic* neutrality — and it was the latter, not the former, on which the holding depended.

Propositions 5–8 are *local* results: they concern the effect of imposing a *small* surcharge starting from none. To obtain a *global* result — comparing no surcharge with *any* level of surcharge — I specialize to a quasi-symmetric linear demand system (a slight generalization of the symmetric demand system in Proposition 4). The general comparative-statics formulas in the proofs of Propositions 5–8 (see the Appendix) involve second-order derivatives of demand and are accordingly complex. With linear demand, all second-order derivatives vanish, yielding clean closed-form expressions. The next proposition shows that while a non-discriminatory tax of size $t$ is "chip-neutral", a non-discriminatory royalty surcharge of size $s$ is not.

**Proposition 9 (Price and Quantity Non-Neutrality of a Surcharge of Any Size).** *Consider a quasi-symmetric linear demand system for differentiated products*[38]

$$q^i(p_1, p_2) = B\left[a_i - p_i + dp_j\right], 0 < d < 1, 0 < a_i, B, i = 1 \; and \; 2.$$

*Then:*

*(i)* *A government tax of size* $t$ *raises both firms' equilibrium all-in modem chip prices by* $\frac{(2+d)}{4-d^2}\, t$ *and reduces quantities by* $B\,\frac{(1-d)(2+d)}{4-d^2}\, t$. *The per-unit tax is chip-neutral in both price and quantity effects.*

*(ii)* *A royalty surcharge of size* $s$ *raises Qualcomm's equilibrium all-in price by* $\frac{3d}{4-d^2}\, s$ *and Intel's by* $\frac{(2+d^2)}{4-d^2}\, s$ *and reduces Qualcomm's output by* $B\,\frac{d(1-d^2)}{4-d^2}\, s$ *and Intel's by* $B\,\frac{2(1-d^2)}{4-d^2}\, s$. *The per-unit surcharge is not chip-neutral: it raises Intel's*

[38] The demand system is called quasi-symmetric because the substitution parameters are both equal to $d$, but the demand intercepts $a_i$ can differ. It arises from the quasi-linear utility $U(q_1, q_2; M) = \sum_{i=1}^{2} A_i\, q_i - \frac{d}{2} Q^2 - \frac{1-d}{2} \sum_{i=1}^{2} q_i^2 + M$ (with $Q \equiv q_1 + q_2$ and M the numeraire), giving inverse demands $p_i = A_i - q_i - dq_j$ and hence $q_i = B\left(a_i - p_i + dp_j\right)$ with $B = \frac{1}{1-d^2}$ and $a_i = A_i - dA_j$. The equality of the ordinary cross-price effects, $\frac{\partial q_i}{\partial p_j} = \frac{\partial q_j}{\partial p_i}$, holds here because the quasi-linear utility removes income effects, so the ordinary and compensated demands coincide. This is the special case of Singh and Vives's (1984) differentiated-duopoly demand in which the own-price effects are equal across the two goods; their formulation lets the two differ, while still allowing the intercepts to differ as here. The symmetric demand of Proposition 4 is in turn the special case $A_1 = A_2$. In their analysis of vertical foreclosure, Ordover, Saloner and Salop (1990) examine a symmetric version of the demand system in Proposition 9.

*all-in price by strictly more than Qualcomm's and reduces Intel's output by strictly more than Qualcomm's.*[39]

*Proof.* Under FRAND, the two firms' best response functions are

$$p_1^F(p_2^F) = \frac{a_1 + c_1 + dr^F + dp_2^F}{2} \qquad (Eq\ 6.1)$$

$$p_2^F(p_1^F) = \frac{a_2 + c_2 + r^F + dp_1^F}{2} \qquad (Eq\ 6.2)$$

so that the FRAND-equilibrium prices and quantities are

$$p_1^{F*} = p_1^{0*} + \frac{3d}{4-d^2} r^F, \qquad p_2^{F*} = p_2^{0*} + \frac{(2+d^2)}{4-d^2} r^F,$$

$$q_1^{F*} = q_1^{0*} - B\frac{d(1-d^2)}{4-d^2} r^F, \qquad q_2^{F*} = q_2^{0*} - B\frac{2(1-d^2)}{4-d^2} r^F$$

where

$$p_i^{0*} \equiv \frac{(2a_i + da_j) + (2c_i + dc_j)}{4-d^2}, \qquad q_i^{0*} \equiv B\frac{[2a_i + da_j] - [(2-d^2)c_i - dc_j]}{4-d^2}$$

A government tax shifts both firms' effective marginal costs by $t$ uniformly, so both all-in prices rise by the same amount:

$$p_1^{G*} = p_1^{F*} + \frac{(2+d)}{4-d^2} t, \qquad p_2^{G*} = p_2^{F*} + \frac{(2+d)}{4-d^2} t$$

As a result, both Qualcomm and Intel lose the same amount of sales, as well:

$$q_1^{G*} = q_1^{F*} - B\frac{(1-d)(2+d)}{4-d^2} t, \qquad q_2^{G*} = q_2^{F*} - B\frac{(1-d)(2+d)}{4-d^2} t$$

[39] Proposition 9 compares no surcharge with a surcharge — or tax — of any size, but its closed forms describe an interior equilibrium and hold only while both firms sell positive quantities. The positive-sales assumption of Section 2 and the Proposition 4 interior bound in footnote 26 supply this restriction locally; the global statement calls for the analogous parameter domain explicitly. Under a surcharge both quantities stay positive if and only if $r^F + s < \frac{(2a_2 + d\,a_1) - (2-d^2)c_2 + d\,c_1}{2(1-d^2)}$ (Intel) and $r^F + s < \frac{(2a_1 + d\,a_2) - (2-d^2)c_1 + d\,c_2}{d(1-d^2)}$ (Qualcomm); under a tax, both $2(1-d^2)\,r^F + (1-d)(2+d)\,t < (2a_2 + d\,a_1) - (2-d^2)c_2 + d\,c_1$ (Intel) and $d(1-d^2)r^F + (1-d)(2+d)t < (2a_1 + d\,a_2) - (2-d^2)c_1 + d\,c_2$ (Qualcomm) must hold. Intel's output falls faster in the policy variable, so in the quasi-symmetric region of interest its constraint is the one that tends to bind first; which constraint binds first more generally depends also on the firms' initial output slack, not on this rate alone. In the symmetric specialization $a_i = (1-d)A$ and $c_i = c$, the surcharge bound collapses to footnote 26's $\frac{2+d}{2(1+d)}(A-c)$. Beyond this region the formulas would return negative Intel sales — a purely algebraic boundary, distinct from the economic exit in Corollary 1, where it is Intel's profit, not its output, that falls below its avoidable fixed cost.

A surcharge of size $s$ shifts $r^F$ to $s + r^F$, raising Intel's price more than Qualcomm's:

$$p_1^{N*} = p_1^{F*} + \frac{3d}{4-d^2}s, \qquad p_2^{N*} = p_2^{F*} + \frac{(2+d^2)}{4-d^2}s$$

Intel's sales decrease by more than Qualcomm's:

$$q_1^{N*} = q_1^{F*} - B\frac{d(1-d^2)}{4-d^2}s, \qquad q_2^{N*} = q_2^{F*} - B\frac{2(1-d^2)}{4-d^2}s \qquad \blacksquare$$

Intel's profits fall on both counts — fewer units and lower margin — confirming Proposition 2 in this parametric example. Qualcomm faces opposing forces: a higher margin offsets some lost sales. Whether the net effect on Qualcomm's profit is positive depends on the size of the surcharge relative to the FRAND royalty, which is the question Proposition 4 addresses.

Figure 1 contrasts the effects on equilibrium prices of a non-discriminatory tax versus a non-discriminatory surcharge. Intel's best-response function shifts upward by the same magnitude when $s = t$. This reflects that for Intel, both the tax and the surcharge of the same size reduce its margin by the same amount and thus exert the same UPP on Intel. In the linear model, Intel's best-response function shifts upward by $\frac{t}{2}$ under the tax (by adding $t$ to $c_2$) and by $\frac{s}{2}$ under the surcharge (by adding $s$ to $r^F$), which are equal to each other when $s = t$. See $p_2^F(p_1^F)$ in (Eq 6.2). In contrast, a royalty surcharge exerts a *lower* UPP on Qualcomm than a tax of equal size so that Qualcomm's best-response function shifts by a *smaller* magnitude to the right under the surcharge than under the tax, because under (A3), $sq_1^2 < -tq_1^1$ when $s = t$. In the linear model, Qualcomm's best-response function shifts to the right by $\frac{t}{2}$ under the tax (by adding $t$ to $c_1$) but only by $d\frac{s}{2}$ (by adding $s$ to $r^F$) under the surcharge. This gives $d\frac{s}{2} < \frac{t}{2}$ when $s = t$ — because the diversion ratio $d$ is less than 1. See $p_1^F(p_2^F)$ in (Eq 6.1). For illustrative purposes, Figure 1 takes $p_1^{F*} = p_2^{F*}$, so that $p_1^{G*} = p_2^{G*}$ but $p_1^{N*} < p_2^{N*}$. The government tax is "chip-neutral," because it raises both prices by equal amounts, but the NLNC royalty surcharge is not, because it raises Intel's price more than Qualcomm's.

Figure 1: Non-discriminatory tax vs. non-discriminatory surcharge

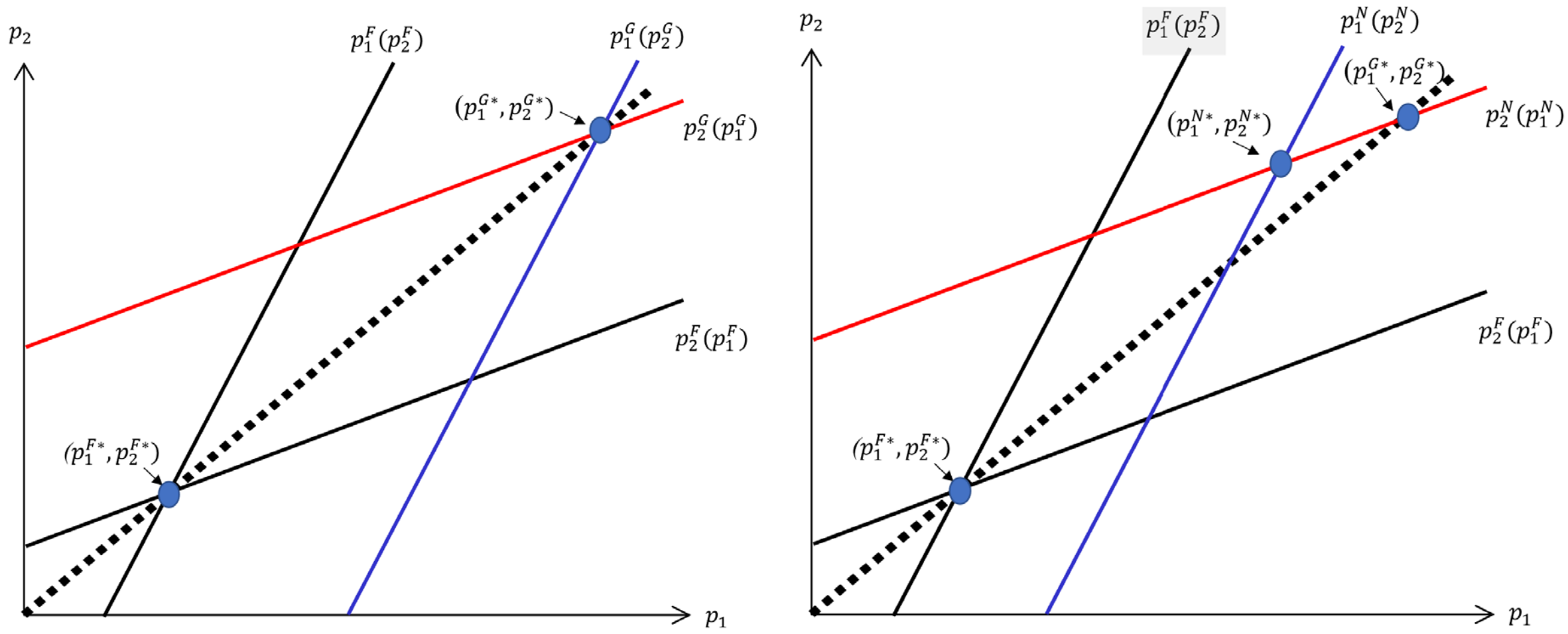

Panel A: Government Tax (chip-neutral) Panel B: NLNC Surcharge (not chip-neutral)

Note: for illustrative purposes, the equilibrium prices under the FRAND royalty are assumed to be equal to each other.

I emphasize the critical role of the Partial Demand Diversion Condition (A3). When the diversion ratio is exactly one, the chip-neutrality argument goes through. For example, consider Hotelling's "linear city" model. Handset makers of type $\omega \in (0,1)$, with unit measure, each demand at most one chip and value firm 1's chip at $v_1(\omega) = w_1 - \alpha_1\omega$ and firm 2's at $v_2(\omega) = w_2 - \alpha_2(1-\omega)$.[40] With $w_i$ large enough that all handset makers are served, the implied demand is

$$q^i(p_1, p_2) = B\big(a_i - p_i + p_j\big), \text{where } B = \frac{1}{\alpha_1 + \alpha_2} \text{ and } a_i = \alpha_j + \big(w_i - w_j\big).$$

This is the demand system in Proposition 9 with $d = 1$ replacing $0 < d < 1$. At $d = 1$, both firms' all-in prices rise by exactly the amount of the surcharge:

$$p_1^{N*} = p_1^{F*} + s, \qquad p_2^{N*} = p_2^{F*} + s.$$

But this reflects a peculiar feature of the Hotelling model: it assumes that total market size is *fixed* — industry output is invariant to prices — so (A3) and (A4) both fail. The model is well suited to other strategic-behavior questions (Whinston 1990 on tying; Choi and Yi 2000 on vertical foreclosure), but ill-suited to testing chip neutrality. Crucially, chip neutrality here does not make the surcharge benign. It remains a naked tax on transactions between Intel and handset makers — though in this fixed-market limit its bite reduces to the surplus transfer described next, not the broader chip-choice and margin effects of Section 5. The fixed-market assumption neutralizes only one specific dimension of harm — the distortion of chip choice. The surcharge still raises both firms' prices, each by exactly $s$. For Intel this is a full pass-through — its price rises by the same $s$ that its cost does — so its margin is unchanged. And because each firm's demand depends only on the difference between the two prices (a second feature of the Hotelling model, alongside fixed market size), and that difference is unchanged when both prices rise by $s$, Intel's output is unchanged too; with margin and output both unchanged, so is its profit. Qualcomm, by contrast, collects the surcharge: its margin rises, and with output unchanged, so does its profit. The burden falls on the handset makers, who pay the higher prices and pass them on, at least in part, to final consumers. So even here, where chip neutrality holds and the rival is untouched, the surcharge is no innocuous tax: it transfers surplus from consumers to Qualcomm.

[40] This version of the Hotelling model slightly generalizes Whinston's (1990) Example 2 (with typo in firm 2's demand corrected so that $\alpha_2$ is multiplied by $(1-\omega)$, not $\omega$).

Shapiro and Waehrer (2023) reach this qualitative harm but not the comparison that decides the chip-neutrality question. Their central argument, that Qualcomm can re-cut its own chip price so that only the royalty on rival handsets matters, establishes that the surcharge burdens the rival; but by treating the charge as effectively rival-only, it does not deliver the equilibrium comparison of the two firms' all-in prices. Where they do touch that comparison (their Section IV.G), they offer two observations that pull in opposite directions: in one, the rival's cost-driven price increase induces Qualcomm to raise its own price, likely by less, which would make the rival's increase the larger; in the other, the surcharge's dampening of Qualcomm's incentive to compete raises Qualcomm's price; they leave the rival's response unstated, but the same best-response logic would have the rival meet it with a smaller increase of its own, making Qualcomm's the larger. The two are never weighed against each other. What is left open is exactly the question the Ninth Circuit found dispositive: do the two all-in prices rise by the same amount, leaving chip choice undistorted, or by different amounts? Propositions 5 to 8 answer it. Under the conditions that would make a non-discriminatory tax chip-neutral, the increases are unequal, the rival's all-in price rises by strictly more than Qualcomm's, and the quantity that settles the comparison is the diversion ratio, with neutrality recovered only in the knife-edge case where it equals one.

## 7. Discriminatory vs. Non-Discriminatory Surcharges

The preceding sections have analyzed a non-discriminatory surcharge applied uniformly to every handset. The actual record is more complicated: Qualcomm provided "chip incentive funds" — discounts tied to the purchase of Qualcomm chips — to a select group of handset makers.[41] These funds reduced the effective per-unit total royalty (the FRAND rate plus the surcharge), but only for handsets containing Qualcomm chips and only for the recipients of the funds.

As Shapiro and Waehrer (2023) observe, however, this discrimination does not change the equilibrium analysis. When Qualcomm can freely adjust its nominal chip price, only the *all-in* price of Qualcomm's chip enters its profit margin (Eq 5.1) — the breakdown into nominal price and surcharge is irrelevant. The anticompetitive bite comes from the surcharge on rival chips, not from whether the surcharge is nominally symmetric across chip suppliers.

[41] *See supra* note 10.

The Shapiro–Waehrer observation leaves open a question central to the antitrust analysis: if discrimination is irrelevant for equilibrium prices, why did Qualcomm offer chip incentive funds at all? The answer lies in the dynamics of royalty negotiation. If a handset maker accepts paying $r^F + s$, Qualcomm can present the agreement as a market-tested benchmark for what FRAND requires — strengthening its hand against any handset maker that resists.[42] Viewed this way, the chip incentive funds are not the legitimate, above-cost discounts Qualcomm portrays them to be.[43] They are a tool for entrenching the market-wide royalty surcharge.

Suppose Qualcomm gives a chip incentive fund of total size $\$D$, tied to purchases of its own chips, to a subset of handset makers accounting for a share $\theta$ of its sales. Qualcomm's profit is then

$$\pi^1(p_1^N, p_2^N; D) = [\theta(p_1^N - c_1)q^1(p_1^N, p_2^N) - D] + (1-\theta)(p_1^N - c_1)q^1(p_1^N, p_2^N) + (r^F + s)q^2(p_1^N, p_2^N) \qquad (Eq\ 7.1)$$

Because the fund is conditional on the recipient's purchasing Qualcomm chips, it is an inducement to remain a Qualcomm customer rather than a no-strings rebate. As a fixed sum, however, it is inframarginal: it lowers a recipient's total outlay without changing the per-unit price it faces at the margin, so the per-unit pricing margin — and with it the equilibrium prices and outputs of the previous sections — is unchanged. The quantity $\$\frac{D}{\theta q^1}$ is the average per-unit value of the fund to a recipient, the gap between the headline rate $r^F + s$ and what it effectively pays; the fund's competitive bite lies in the negotiation dynamics noted above, not in the marginal pricing equilibrium. These negotiation dynamics are intertemporal and lie outside the static model developed here; I treat them informally and leave their formalization to future work.

A different pattern appeared in the first Korean Qualcomm case:[44] Qualcomm charged explicitly lower royalty rates on chips installed in handsets exported by Korean handset makers.[45] To capture this, suppose Qualcomm gives a per-unit royalty discount of $x$ to a subset of handset makers accounting for a share $\theta$ of its chip sales, with $\theta$ taken as exogenous. Assume Qualcomm cannot price-discriminate on the nominal chip price. (Without this restriction, Qualcomm would offset the royalty discount through a higher chip price.) Qualcomm's profit is

[42] Opening Brief for Appellant Qualcomm at 31 ("Qualcomm's established royalty is the best measure of what constitutes a reasonable royalty."), *FTC v. Qualcomm Inc.*, 969 F.3d 974 (9th Cir. 2020) (No. 19-16122), Dkt. Entry 77-2 (Aug. 23, 2019).
[43] Reply Brief for Appellant Qualcomm at 49, *supra* note 30 ("The District Court erred in its sweeping condemnation of discounts that Qualcomm has provided—or merely offered—to its modem chip customers. Antitrust law favors discounts because they reflect price competition and lower prices to consumers. Accordingly, above-cost discounts are generally considered *per se* legal.").
[44] Korea Fair Trade Commission Decision, Case No. 2009JiSik0329.
[45] Yi and Kim (2018), at 319–20.

$$\pi^1(p_1^N, p_2^N; x) = \theta(p_1^N - x - c_1)q^1(p_1^N, p_2^N) + (1-\theta)(p_1^N - c_1)q^1(p_1^N, p_2^N) + (r^F + s)q^2(p_1^N, p_2^N)$$

$$= (p_1^N - c_1 - \theta x)q^1(p_1^N, p_2^N) + (r^F + s)q^2(p_1^N, p_2^N) \qquad (Eq\ 7.2)$$

Because this discount applies only to handsets containing Qualcomm's own chips, it does not affect the royalty Qualcomm collects on handsets containing rival chips, so that term is unchanged. The only difference from Section 5 is that Qualcomm's effective marginal cost is raised by the expected discount $\theta x$. Results on the anticompetitive effects of the surcharge carry over.[46] Equation 7.2 evaluates both groups at the common all-in price and nets the discount off Qualcomm's average margin. A fully disaggregated treatment would place recipients on $q^1(p_1^N - x,\ p_2^N)$, since their lower royalty lowers their all-in cost; that reallocates output among Qualcomm's own buyers but leaves the rival-chip term $(r^F + s)q^2$ — which carries the surcharge's non-neutrality — untouched, so the anticompetitive conclusions are unaffected. Section 7 is thus interpretive rather than load-bearing: the static models confirm that discrimination leaves the anticompetitive conclusions of Sections 5–6 intact, so the substance of the chip-incentive funds lies in the dynamics of royalty negotiation, not the static comparison.

## 8. Ad-Valorem Royalty Surcharges on Handset Prices

Sections 3–7 analyzed Qualcomm's royalty as a per-handset charge — the form it took for a substantial part of the market. Apple, a large buyer, paid a flat $7.50 per handset; and for the other makers the percentage royalty was capped at $20, so their handsets priced above $400 carried a flat $20 (Section 3). Genuine ad-valorem pricing — typically 5% of the handset price — applied to the remainder: non-Apple handsets priced under $400. It is that case I take up here. The distinction matters for the analysis: a per-handset surcharge, under fixed proportions (one chip per handset), is collected on the number of chips sold, so one can work directly with the derived demand for chips. An ad-valorem surcharge is instead levied on the handset maker's revenue — price times quantity — so the analysis must also track how modem chip prices pass through to handset prices. I develop the ad-valorem analysis formally in Appendix B and summarize its findings here.

The ad-valorem case is a parallel formal treatment, not an illustration: it carries the refutation to Qualcomm's neutrality premise itself. Its price and output effects hold under the same weak pass-through conditions, (A6)–(A7); the surcharge-versus-tax and chip-choice results require the stronger downstream-revenue diversion condition (A7′), and only the benchmark results specialize to the quasi-symmetric linear derived demand and constant symmetric pass-through that Proposition 9 already invokes, with the zero-FRAND knife-edge examined in a deliberately simplified two-maker downstream model. The formal

[46] The only exception is Proposition 4, which relies on $c_1 = c_2 = c$ to obtain a clean condition. Proposition 4 extends straightforwardly to the case where $c_1 \neq c_2$, at the cost of messier notation.

development — including the downstream-revenue diversion condition and the measure-zero offset of Proposition B9 — is in Appendix B.

The anticompetitive conclusions are unchanged. Under weak assumptions on the pass-through rates ((A6) and (A7)), an ad-valorem royalty surcharge behaves like an ad-valorem downstream tax: it raises both modem chip prices and both handset prices, and reduces total output, with no offsetting efficiency in the model (Propositions B1 and B2). Whether the surcharge is levied per handset or on handset value is therefore irrelevant to its anticompetitive character.

The chip-neutrality question requires more care. With per-handset royalties, the Partial (Derived) Demand Diversion Condition (A3) was enough to show that Qualcomm has a weaker incentive to raise its price under the surcharge than under an equal tax. With ad-valorem royalties this no longer suffices; the relevant condition is the Partial Downstream Revenue Diversion Condition (A7′), the downstream-revenue analog of (A3): when one modem chip maker raises its price, the fall in revenue on handsets carrying its chip is only partly offset by the rise in revenue on handsets carrying the rival's. Under (A7′), the surcharge–tax comparison parallels Proposition 5 — both firms' prices rise less under the surcharge than under an equal tax, but the gap is larger for Qualcomm (Proposition B3). Under an ad-valorem counterpart of Qualcomm's Implicit Assumption, the comparison with no surcharge parallels Propositions 6–8: a small surcharge raises Intel's price by more than Qualcomm's (Proposition B4), reduces Intel's output (Proposition B5), and — if Qualcomm's output falls — reduces Intel's by strictly more (Proposition B6).

The profit results carry over once demand is specialized. When the derived demand for chips is quasi-symmetric and linear (as in Proposition 9) and the pass-through rates are constant and symmetric, the surcharge lowers Intel's margin and—under further conditions on the pass-through rates—its profit (Proposition B7, paralleling part (ii) of Proposition 2), and, provided the total ad-valorem royalty is small relative to Intel's gross upstream margin (and under further conditions on the pass-through rates), it raises Qualcomm's profit (Proposition B8, paralleling Proposition 3).

The decisive break with the per-handset case concerns the chip-neutrality benchmark itself. With per-handset royalties, a non-discriminatory tax is chip-neutral whenever the derived demand is linear with symmetric price effects (Proposition 9); Qualcomm's Implicit Assumption holds, so its analogy between the surcharge and a neutral tax is at least well-posed — and the analogy then fails, because the surcharge raises Intel's price by more. With ad-valorem royalties the benchmark itself collapses: even under full symmetry — symmetric linear derived demand, constant symmetric pass-through, and equal chip costs — a non-discriminatory tax is chip-neutral if and only if the FRAND royalty rate is zero (Proposition B9). Because it was undisputed in *FTC v. Qualcomm* that the FRAND royalty rate was positive, the premise on which the Ninth Circuit's holding rests fails in the ad-valorem case as well. Two distinct failures therefore underlie the chip-neutrality holding: for per-handset royalties the surcharge is not equivalent to a tax even where the tax is neutral, while for ad-valorem royalties no neutral tax exists once the FRAND rate is positive, save for the measure-zero offset identified in Proposition B9.

Rather than stop there, I examine the knife-edge case in which the premise does hold — a zero FRAND rate under full symmetry. In a deliberately simplified downstream model with two handset makers, Samsung (buying only Qualcomm chips) and Apple (buying only Intel chips), a small surcharge still raises Intel's price by more whenever the downstream diversion ratio between the two handsets lies below a threshold of about 0.7, and even above that threshold whenever modem chip costs are sufficiently high (Proposition B10; the threshold corresponds to an industry-wide pass-through rate of about 0.77 or an upstream diversion ratio of about 0.46). Reported diversion ratios between Apple and Samsung handsets lie well below these thresholds, so even in this most favorable case for Qualcomm the surcharge is not chip-neutral.

## 9. Concluding remarks

In this paper, I have examined Qualcomm's chip-neutrality argument, which the Ninth Circuit accepted. The argument trades on two senses of neutrality. A charge is neutral on its face when it does not discriminate by chip source — when, like a tax, it falls on every handset at the same rate whatever modem chip it carries. A charge is neutral in economic effect when it leaves the relative all-in price of the two chips, and so the handset maker's chip choice, undisturbed. Qualcomm turned the FTC's tax analogy into a defense by sliding from the first to the second: because the surcharge is facially neutral, it argued, the playing field between Qualcomm and its rivals stays level.[47] Stated as the syllogism it implicitly is, the defense runs:

| P1. The surcharge is a non-discriminatory charge on all handsets, like a tax. |
|---|
| P2. A non-discriminatory tax is neutral across chip suppliers. |
| ∴ The surcharge is neutral across chip suppliers. |

P2 is never stated, yet it is what lends the defense its force, since taxes are presumed neutral, and I have called it Qualcomm's Implicit Assumption. The per-handset and ad-valorem royalties fail the syllogism at different premises.

For a per-handset royalty, grant P2 in its strongest form, the quasi-symmetric linear demand under which a non-discriminatory tax is exactly neutral (Proposition 9). P1 still fails. What fails is not the facial claim that the surcharge is a non-discriminatory charge on all handsets, which I grant, but the "like a tax" clause appended to it. A tax and the surcharge bear on the rival's pricing identically and differ only in their effect on Qualcomm, which collects the surcharge rather than remitting it: where a tax lets Qualcomm bank the levy it avoids on its own lost sales when it raises its price, the surcharge returns only the royalty on the sales it diverts to the rival, and because the diversion ratio is below one (Assumption (A3)) that pickup is the smaller of the two. Qualcomm's incentive to raise its all-in price is therefore weaker under the surcharge than under the tax, the rival's all-in price rises by more than Qualcomm's, and handset makers are pushed toward Qualcomm (Propositions 5 to 8). On the very terrain where a tax is neutral, the surcharge is not.

[47] *See* the quote from Qualcomm's Motion to Dismiss below.

For an ad-valorem royalty, the form a large part of Qualcomm's royalties took, the defense fails one step earlier, at P2. Even under full symmetry, a non-discriminatory ad-valorem tax is itself neutral across chip suppliers only if the FRAND rate is zero (Proposition B9); because that rate was undisputedly positive, there is generically no neutral tax for the surcharge to resemble — save for the measure-zero set of offsetting asymmetric primitives that Proposition B9 preserves — and the analogy collapses before the surcharge is introduced. This sharpens rather than disputes Qualcomm's entitlement: the refutation never questions Qualcomm's right to a reasonable royalty, yet under an ad-valorem royalty even the entitled FRAND component is enough to break the neutrality of a non-discriminatory tax.

Chip neutrality was not a new theory Qualcomm devised on appeal. It was Qualcomm's central argument from the outset of *FTC v. Qualcomm*. In its April 3, 2017 motion to dismiss the FTC's January 17, 2017 complaint, Qualcomm argued:

> "The alleged effects of this purported "tax" are to diminish demand for competitors' chips and reduce competitors' margins. But the chip-neutral nature of Qualcomm's royalties—the fact that the royalties do not change based on the source of the modem chip—means that they do not and cannot create any incentive or disincentive for the handset maker to purchase any particular firm's modem chip. From a customer's perspective, the playing field between Qualcomm and its rivals is level for each modem chip sale. Whether the royalties are "elevated" or not, customers face no discriminatory penalty (or "tax") for the purchase of competitors' modem chips." [48]

The district court rejected the motion.[49] At the January 2019 trial, Qualcomm's expert again emphasized the "chip-neutral" character of Qualcomm's royalty.[50] The district court's May 2019 Findings of Fact and Conclusions of Law again sided with the FTC.[51]

---

[48] Defendant Qualcomm Incorporated's Motion to Dismiss at 1, *FTC v. Qualcomm Inc.*, No. 17-cv-00220-LHK (N.D. Cal. Apr. 13, 2017), ECF No. 69 (footnote omitted) (unredacted version at ECF No. 143).

[49] Order Denying Motion to Dismiss, *FTC v. Qualcomm Inc.*, No. 17-cv-00220-LHK (N.D. Cal. June 26, 2017), ECF No. 133.

[50] Trial Transcript at 1891, *FTC v. Qualcomm Inc.*, No. 17-cv-00220-LHK (N.D. Cal. Jan. 25, 2019) (testimony of Aviv Nevo)

> ("[T]he royalty is the same royalty regardless of which chip the OEM uses, and in that sense, it cancels out, if you will, the decision of which chip to use. … [I]f I'm an OEM, I'm thinking I'm going to use chip A or chip B, either way, the royalty is being paid. So it should not affect the decision of which chip to use. So in that sense I call it chip neutral.").

[51] *Qualcomm*, 411 F. Supp. 3d at 791

> ("Qualcomm's unreasonably high royalty rates enable Qualcomm to control rivals' prices because Qualcomm receives the royalty even when an OEM uses one of Qualcomm's rival's chips. Thus, the 'all-in' price of any modem chip sold by one of Qualcomm's rivals effectively includes two components: (1) the nominal chip price; and (2) Qualcomm's royalty

Qualcomm raised the same argument on appeal, without new substance:

> "Qualcomm's royalties are "chip-neutral"; chip suppliers compete for business on an even playing field."[52]

This time, the appellate panel sided with Qualcomm. The natural question is why. A notable development on appeal was the intervention of the Antitrust Division of the Department of Justice — the FTC's sister antitrust agency — as amicus on Qualcomm's behalf. The first Trump Administration's Antitrust Division argued that "the district court erroneously sanctioned Qualcomm under the Sherman Act for charging high royalties, injury merely to customers, and violating patent law."[53] The premise that the FTC equated high royalties with competitive harm is wrong: the anticompetitive effects of NLNC arise from the surcharge on rivals, as Shapiro and Waehrer (2023) note and as this paper demonstrates formally. The Antitrust Division also endorsed the chip-neutrality theory, asserting that "OEMs pay for use of Qualcomm's SEPs that are essential to every cellular device produced, *regardless of which supplier's chip is used*."[54]

Generalist appellate courts often struggle with antitrust cases that turn on subtle economic reasoning, and one might be tempted to read the Ninth Circuit's chip-neutrality holding as an unremarkable instance of that difficulty. What sets *FTC v. Qualcomm* apart is the intervention of the DOJ Antitrust Division as amicus on Qualcomm's behalf. Shapiro and Waehrer (2023) treat that intervention as one of several factors contributing to reversal and call for additional economics training for appellate judges. The argument of this paper points to a more specific concern. When the federal government's two expert antitrust agencies take opposing positions on a question of competitive harm, the appellate court is no longer asked to weigh agency expertise against unaided generalist judgment; it is asked to choose between two specialist views. The propositions established here show that, on the economics, the choice was not a close call. A companion paper examines the Antitrust Division's reasoning, and its errors, in detail.

---

> surcharge. To Qualcomm, the surcharge represents 'higher profits,' both because the surcharge brings additional revenue to Qualcomm, and 'because the reduction in competition enable[s]' Qualcomm 'to capture more of the [modem chip] market.'").

Unlike the FTC, the district court excludes the FRAND royalty from the all-in price; this does not affect the analysis.

[52] Reply Brief for Appellant Qualcomm at 4, *supra* note 30.

[53] Brief of the United States of America as Amicus Curiae in Support of Appellant and Vacatur at i, *FTC v. Qualcomm Inc.*, 969 F.3d 974 (9th Cir. 2020) (No. 19-16122), Dkt. Entry 86 (Aug. 30, 2019).

[54] *Id*. at 18 (original emphasis).

**References**

**Bulow, Jeremy I., and Paul Pfleiderer**. 1983. "A Note on the Effect of Cost Changes on Prices." *Journal of Political Economy* 91(1): 182-185.

**Choi, Jay Pil, and Sang-Seung Yi.** 2000. "Vertical Foreclosure with the Choice of Input Specifications." *RAND Journal of Economics* 31(4): 717–743.

**Dickson, Vaughan.** 2010. "Cost Pass-Through Elasticities, Concentration and Productivity Growth." *Applied Economics Letters* 17(7): 663–66.

**Fan, Ying, and Chenyu Yang. 2020**. "Competition, Product Proliferation, and Welfare: A Study of the US Smartphone Market." *American Economic Journal: Microeconomics* 12(2): 99–134.

**Farrell, Joseph, and Carl Shapiro.** 2010. "Recapture, Pass-Through, and Market Definition." *Antitrust Law Journal*, 76(3): 585–604.

**Grzybowski, Lukasz, and Ambre Nicolle**. 2021. "Estimating Consumer Inertia in Repeated Choices of Smartphones." *Journal of Industrial Economics*. 69(1): 33–82.

**Goldberg, Pinelopi Koujianou, and Michael M. Knetter.** 1997. "Goods Prices and Exchange Rates: What Have We Learned?" *Journal of Economic Literature,* 35(3): 1243–72.

**Krattenmaker, Thomas G., and Steven C. Salop.** 1986. "Anticompetitive Exclusion: Raising Rivals' Costs to Achieve Power Over Price." *Yale Law Journal* 96(2): 209–293.

**Layne-Farrar, Anne, Gerard Llobet, and Jorge Padilla.** 2014. "Patent Licensing in Vertically Disaggregated Industries: The Royalty Allocation Neutrality Principle." *Communications & Strategies* 95(3): 61–84.

**Llobet, Gerard, and Damien Neven.** 2023. "Investment and Patent Licensing in the Value Chain." *Journal of Competition Law & Economics* 19(4): 527–555.

**Llobet, Gerard, and Jorge Padilla.** 2016. "The Optimal Scope of the Royalty Base in Patent Licensing." *Journal of Law and Economics* 59(1): 45–73.

**Ordover, Janusz A., Garth Saloner, and Steven C. Salop**. 1990. "Equilibrium Vertical Foreclosure." *American Economic Review,* 80(1): 127–42.

**Padilla, Jorge, and Koren W. Wong-Ervin.** 2017. "Portfolio Licensing to Makers of Downstream End-User Devices: Analyzing Refusals to License FRAND-Assured Standard-Essential Patents at the Component Level." *The Antitrust Bulletin* 62(3): 494–513.

**Peitz, Martin, and Markus Reisinger.** 2014. "Indirect Taxation in Vertical Oligopoly." *Journal of Industrial Economics* 62 (4): 709–55.

**Roberson, Michael**. 2016. "Price Elasticity for Smartphones in the United States: Results from Three Methodological Approaches." PhD diss., Texas Tech University.

**Seade, Jesus.** 1985. "Profitable Cost Increases and the Shifting of Taxation: Equilibrium Response of Markets in Oligopoly." Http://ideas.repec.org/p/wrk/warwec/260.html.

**Shapiro, Carl.** 1996. "Mergers with Differentiated Products." *Antitrust*, 10(2): 23-30.

**Shapiro, Carl, and Keith Waehrer.** 2023. "Using and Misusing Microeconomics: Federal Trade Commission v. Qualcomm." In *Antitrust Economics at a Time of Upheaval: Recent Competition Policy Cases on Two Continents*, edited by John Kwoka, Tommaso Valletti, and Lawrence J. White. Boston: Competition Policy International.

**Singh, Nirvikar, and Xavier Vives.** 1984. "Price and Quantity Competition in a Differentiated Duopoly." *RAND Journal of Economics* 15(4): 546–554.

**U.S. Department of Justice & FTC**. 2023. Merger Guidelines.

**Weyl, E. Glen, and Michal Fabinger.** 2013. "Pass-through as an Economic Tool: Principles of Incidence under Imperfect Competition." *Journal of Political Economy* 121(3): 528−583.

**Whinston, Michael D.** 1990. "Tying, Foreclosure, and Exclusion." *American Economic Review* 80(4): 837–859.

**Yi, Sang-Seung, and Yoonhee Kim.** 2018. "FRAND in Korea." In *The Cambridge Handbook of Technical Standardization Law: Competition, Antitrust, and Patents*, edited by Jorge L. Contreras. Cambridge: Cambridge University Press.

## Appendix A: Proofs

Under (A5), the best-response functions slope upward with slopes less than 1: $0 < \frac{dp_i^G(p_j^G)}{dp_j^G} = -\frac{\pi_{ij}^i(p_1^G,p_2^G)}{\pi_{ii}^i(p_1^G,p_2^G)} < 1$. When (A5) is replaced with (A5′), the slopes are less than ½.

**Lemma 1.** *Under (A*1*), (A*2*), (A*4*)–(A5′), the best-response functions in the FRAND and tax cases slope upward with slopes less than* 1/2.

*Proof of Lemma* 1. Since $q_1^1 < -q_2^1 < 0$, $q_2^2 < -q_1^2 < 0$, $q_{11}^1 \le -2q_{12}^1 \le 0, q_{22}^2 \le -2q_{12}^2 \le 0$, and $q_{11}^2 \le -2q_{12}^2 \le 0$, we have

$$0 < \frac{dp_1^G(p_2^G)}{dp_2^G} = -\frac{\pi_{12}^1(p_1^G,p_2^G)}{\pi_{11}^1(p_1^G,p_2^G)} = -\frac{q_2^1 + (p_1^G - c_1 - t)q_{12}^1 + r^F q_{12}^2}{2q_1^1 + (p_1^G - c_1 - t)q_{11}^1 + r^F q_{11}^2} < \frac{1}{2}$$

$$0 < \frac{dp_2^G(p_1^G)}{dp_1^G} = -\frac{\pi_{12}^2(p_1^G,p_2^G)}{\pi_{22}^2(p_1^G,p_2^G)} = -\frac{q_1^2 + (p_2^G - c_2 - r^F - t)q_{12}^2}{2q_2^2 + (p_2^G - c_2 - r^F - t)q_{22}^2} < \frac{1}{2}$$

One obtains the FRAND case by setting $t = 0$. ■

*Proof of Proposition* 1.

Total differentiation of the two firms' first-order conditions yields

$$\begin{bmatrix} \pi_{11}^1 & \pi_{12}^1 \\ \pi_{12}^2 & \pi_{22}^2 \end{bmatrix} \begin{bmatrix} \frac{dp_1^G}{dt} \\ \frac{dp_2^G}{dt} \end{bmatrix} = -\begin{bmatrix} \pi_{1t}^1 \\ \pi_{2t}^2 \end{bmatrix} = \begin{bmatrix} q_1^1(p_1^G,p_2^G) \\ q_2^2(p_1^G,p_2^G) \end{bmatrix}$$

Applying Cramer's rule and dividing numerator and denominator by $\pi_{11}^1\pi_{22}^2$ yields[55]

$$\frac{dp_1^{G*}}{dt} = \frac{\left[\frac{q_1^1}{\pi_{11}^1} + \frac{q_2^2}{\pi_{22}^2}\left(\frac{dp_1^G(p_2^G)}{dp_2^G}\right)\right]}{\left[1 - \left(\frac{dp_1^G(p_2^G)}{dp_2^G}\right)\left(\frac{dp_2^G(p_1^G)}{dp_1^G}\right)\right]} > 0$$

under (A1) and (A5). Similarly, $\frac{dp_2^{G*}}{dt} > 0$. Since both prices increase, total output decreases under (A3) and (A4). To see why, decompose the increase in modem chip prices in two steps. At this level of generality, I cannot determine which price rises more. Suppose that firm *i*'s price rises more. First, raise both prices by the increase in firm *j*'s price, $j \neq i$. Under (A4), each firm sells fewer modem chips. Second, increase firm *i*'s price to the new equilibrium level. Under (A3), the increase in firm *j*'s output falls short of the reduction in firm *i*'s output. In sum, fewer firm *i*'s modem chips are sold. While I cannot rule out the possibility that firm *j* sells more modem chips, total output of modem chips decreases.

[55] The first term in the numerator can be interpreted as the direct effect of an increase in the tax on Qualcomm's price, and the second term as the indirect effect operating through Intel's price response.

I now move to the proof of part (ii). When (A5′) is added, $0 < \frac{dp_1^G(p_2^G)}{dp_2^G}, \frac{dp_1^G(p_2^G)}{dp_2^G} < \frac{1}{2}$ as shown in Lemma 1, so that the denominator of $\frac{dp_1^{G*}}{dt}$ is *strictly* greater than ¾. And $0 < \frac{q_i^i}{\pi_{ii}^i} \leq \frac{1}{2}, i = 1, 2$ under (A1) and (A5′) so that the numerator is *strictly* less than 3/4. As a result, $0 < \frac{dp_1^{G*}}{dt} < 1$.[56] The same argument applied to firm 2 yields $0 < \frac{dp_2^{G*}}{dt} < 1$.

The effect of the tax on Intel's equilibrium profit is

$$\frac{d\pi^2(p_1^{G*}, p_2^{G*})}{dt} = \frac{\partial \pi^2}{\partial t} + \frac{\partial \pi^2}{\partial p_1}\frac{dp_1^{G*}}{dt} = -q^2 + [(p_2^{G*} - c_2 - r^F - t)q_1^2]\frac{dp_1^{G*}}{dt}$$

$$= -q^2 + \left[-\frac{q^2}{q_2^2}q_1^2\right]\frac{dp_1^{G*}}{dt} = -q^2\left[1 + \frac{q_1^2}{q_2^2}\frac{dp_1^{G*}}{dt}\right] < 0.$$

The first equality applies the envelope theorem; the third uses Intel's FOC. The final expression is negative, since (i) $-1 < \frac{q_1^2}{q_2^2} < 0$ by (A1), (A2), (A4) and (ii) $0 < \frac{dp_1^{G*}}{dt} < 1$ as shown above.

Qualcomm's profit derivative is more involved because of the royalty income but follows the same template:

$$\frac{d\pi^1(p_1^{G*}, p_2^{G*})}{dt} = \frac{\partial \pi^1}{\partial t} + \frac{\partial \pi^1}{\partial p_2}\frac{dp_2^{G*}}{dt} = -q^1 + [(p_1^{G*} - c_1 - t)q_2^1 + r^F q_2^2]\frac{dp_2^{G*}}{dt}$$

$$= -q^1 + [-\frac{q^1 + r^F q_1^2}{q_1^1}q_2^1 + r^F q_2^2]\frac{dp_2^{G*}}{dt}$$

$$= -q^1\left[1 + \frac{q_2^1}{q_1^1}\frac{dp_2^{G*}}{dt}\right] + r^F q_2^2(-\frac{q_1^2 q_2^1}{q_1^1 q_2^2} + 1)\frac{dp_2^{G*}}{dt} < 0$$

The first equality applies the envelope theorem; the third uses Qualcomm's FOC. The first term in the final expression, a mirror image of that for Intel, is negative. The second term is negative because (i) $q_2^2 < 0$,(ii) $-1 < -\frac{q_1^2 q_2^1}{q_1^1 q_2^2} < 0$ by (A1), (A2), and (A4), and (iii) $0 < \frac{dp_2^{G*}}{dt} < 1$. ■

---

[56] If Qualcomm is a monopolist ($q^2 = 0$), it passes through at most 50% of the tax: $\frac{dp_1^G}{dt} = \frac{q_1^1}{\pi_{11}^1} = \frac{q_1^1}{2q_1^1 + (p_1^G - c_1 - t)q_{11}^1} \leq \frac{1}{2}$ under (A1) and (A5′).

The next lemma is the surcharge-case analog of Lemma 1.

**Lemma 2.** *Under (A*1*), (A*2*), (A*4*)–(A*5*′), the best-response functions in the surcharge (NLNC) case slope upward with slopes less than* 1/2.

*Proof of Lemma* 2. Simply replace $r^F$ with $r^F + s$ and set $t = 0$ in the proof of Lemma 1. ■

*Proof of Proposition* 2. Total differentiation of the FOCs (Eq 5.3) and (Eq 5.4) yields

$$\begin{bmatrix} \pi_{11}^1 & \pi_{12}^1 \\ \pi_{12}^2 & \pi_{22}^2 \end{bmatrix} \begin{bmatrix} \frac{dp_1^N}{ds} \\ \frac{dp_2^N}{ds} \end{bmatrix} = \begin{bmatrix} -q_1^2 \\ q_2^2 \end{bmatrix}$$

The only difference from Proposition 1 is that for Qualcomm, $q_1^1$ is replaced with $-q_1^2$ in the right-hand side. Since these two terms have the same positive sign, the effects of the surcharge on Qualcomm's all-in prices are qualitatively similar to those of the tax:

$$\frac{dp_1^{N*}}{ds} = \frac{\left[\frac{-q_1^2}{q_1^1}\frac{q_1^1}{\pi_{11}^1} + \frac{q_2^2}{\pi_{22}^2}\left(\frac{dp_1^N(p_2^N)}{dp_2^N}\right)\right]}{\left[1 - \left(\frac{dp_1^N(p_2^N)}{dp_2^N}\right)\left(\frac{dp_2^N(p_1^N)}{dp_1^N}\right)\right]} > 0$$

under (A1), (A2) and (A5). Similarly, $\frac{dp_2^{N*}}{ds} > 0$. Since both prices increase, total output decreases under (A3) and (A4) for the same reason as in Proposition 1. Qualcomm's margin $p_1^{N*} - c_1$ rises by the full amount of the price increase, since Qualcomm collects the surcharge from itself.

I now move to the proof of part (ii). When (A5′) is added, the four ingredients are bounded as in the proof of Proposition 1: since $0 < \frac{dp_1^N(p_2^N)}{dp_2^N}, \frac{dp_2^N(p_1^N)}{dp_1^N} < \frac{1}{2}$ as shown in Lemma 2, the denominator is strictly greater than ¾. And $0 < \frac{q_1^1}{\pi_{11}^1}, \frac{q_2^2}{\pi_{22}^2} \leq \frac{1}{2}$ under (A1) and (A5′). The difference from Proposition 1's formula is the extra factor $\frac{-q_1^2}{q_1^1}$ multiplying the first numerator term: under (A1)–(A3), this factor lies in $(0,1)$, so that the numerator is again strictly less than ¾, yielding $0 < \frac{dp_1^{N*}}{ds} < 1$. Similarly,

$$\frac{dp_2^{N*}}{ds} = \frac{\left[\frac{q_2^2}{\pi_{22}^2} + \frac{-q_1^2}{q_1^1}\frac{q_1^1}{\pi_{11}^1}\left(\frac{dp_2^N(p_1^N)}{dp_1^N}\right)\right]}{\left[1 - \left(\frac{dp_1^N(p_2^N)}{dp_2^N}\right)\left(\frac{dp_2^N(p_1^N)}{dp_1^N}\right)\right]}$$

The same argument gives $0 < \frac{dp_2^{N*}}{ds} < 1$.

As under the tax, Intel's margin $p_2^{N*} - c_2 - r^F - s$ falls. The envelope-theorem calculation for Intel's profit is structurally the same as in the tax case:

$$\frac{d\pi^2(p_1^{N*}, p_2^{N*})}{ds} = \frac{\partial \pi^2}{\partial s} + \frac{\partial \pi^2}{\partial p_1}\frac{dp_1^N}{ds} = -q^2\left[1 + \frac{q_1^2}{q_2^2}\frac{dp_1^{N*}}{ds}\right] < 0. \qquad \blacksquare$$

*Proof of Proposition* 3. Differentiating Qualcomm's equilibrium profit with respect to $s$ gives

$$\frac{d\pi^1(p_1^{N*}, p_2^{N*})}{ds} = \frac{\partial \pi^1}{\partial s} + \frac{\partial \pi^1}{\partial p_2}\frac{dp_2^N}{ds}$$

$$= q^2 + (r^F + s)q_2^2\frac{dp_2^N}{ds} + (p_1^{N*} - c_1)q_2^1\frac{dp_2^N}{ds}$$

Qualcomm faces the standard royalty-collector trade-off: a one-dollar increase in the surcharge enables Qualcomm to earn $\$q^2$ on infra-marginal units, but reduces revenue by $\$(r^F + s)$ on each unit Intel no longer sells (since $\frac{dp_2^N}{ds} > 0$ and $q_2^2 < 0$). The lost revenue includes both the FRAND royalty *and* the surcharge.

A second, positive effect operates through chip competition. The surcharge raises Intel's all-in price ($\frac{dp_2^N}{ds} > 0$), which diverts demand to Qualcomm's chips ($q_2^1 > 0$). Since Qualcomm earns a positive margin $p_1^{N*} - c_1 > 0$ on those additional sales, this is the business-stealing effect.

Substituting (Eq 5.4) into the second term we obtain

$$= -(p_2^{N*} - c_2 - r^F - s)q_2^2 + (r^F + s)q_2^2\frac{dp_2^N}{ds} + (p_1^{N*} - c_1)q_2^1\frac{dp_2^N}{ds}$$

$$= -q_2^2\left[(p_2^{N*} - c_2) - (r^F + s)\left[1 + \frac{dp_2^N}{ds}\right]\right] + (p_1^{N*} - c_1)q_2^1\frac{dp_2^N}{ds}$$

$$= q_2^2\left[1 + \frac{dp_2^N}{ds}\right](r^F + s) - (p_2^{N*} - c_2)q_2^2 + (p_1^{N*} - c_1)q_2^1\frac{dp_2^N}{ds} > 0$$

which holds if and only if

$$r^F + s < \frac{-q_2^2(p_2^{N*} - c_2) + (p_1^{N*} - c_1)q_2^1\frac{dp_2^N}{ds}}{-q_2^2\left[1 + \frac{dp_2^N}{ds}\right]} \qquad (Eq\ 5.5)$$

Since $0 < \frac{dp_2^N}{ds} < 1$, a *sufficient* condition is

$$r^F + s < \frac{p_2^{N*} - c_2}{2}$$

The bound in $r^F + s < \frac{p_2^{N*} - c_2}{2}$ ignores the business-stealing effect. To see how that effect changes the analysis, suppose the diversion ratio is close to 1, so that $\frac{q_2^1}{-q_2^2} \approx 1$. Then

$$-(p_2^{N*} - c_2 - r^F - s)q_2^2 + (r^F + s)q_2^2 \frac{dp_2^N}{ds} + (p_1^{N*} - c_1)q_2^1 \frac{dp_2^N}{ds}$$

$$= -q_2^2 \left[ (p_2^{N*} - c_2 - r^F - s) + \left[ (p_1^{N*} - c_1)\frac{q_2^1}{-q_2^2} - (r^F + s) \right] \frac{dp_2^N}{ds} \right]$$

$$\approx -q_2^2 \left[ (p_2^{N*} - c_2 - r^F - s) + [p_1^{N*} - c_1 - r^F - s] \frac{dp_2^N}{ds} \right]$$

Intel's FOC gives $p_2^{N*} - c_2 - r^F - s > 0$. If $r^F + s < p_1^{N*} - c_1$, both bracketed terms are positive and the result follows. If $r^F + s > p_1^{N*} - c_1$, the second term is negative, but the bound $0 < \frac{dp_2^N}{ds} < 1$ ensures the bracket remains positive provided $r^F + s < \frac{(p_1^{N*} - c_1) + (p_2^{N*} - c_2)}{2}$. ■

*Proof of Proposition* 4. Proposition 4 is the symmetric special case of the linear demand system in Proposition 9. Imposing the symmetry and parameter restrictions ($a_1 = a_2 = (1-d)A$, and $c_1 = c_2 = c$; $B = \frac{1}{1-d^2}$) and writing $a \equiv a_1 = a_2 = (1-d)A$ for the common intercept, we obtain:

$$p_1^{N*} = p_1^{F*} + \frac{3ds}{4-d^2}, \qquad p_2^{N*} = p_2^{F*} + \frac{(2+d^2)s}{4-d^2}$$

$$q_1^{N*} = q_1^{F*} - \frac{d}{4-d^2}s, \qquad q_2^{N*} = q_2^{F*} - \frac{2}{4-d^2}s$$

where

$$p_1^{F*} = \frac{a+c}{2-d} + \frac{3d}{4-d^2}r^F, \qquad p_2^{F*} = \frac{a+c}{2-d} + \frac{(2+d^2)}{4-d^2}r^F$$

$$q_1^{F*} = \frac{1}{1+d}\left[\frac{(A-c)}{2-d} - \frac{d(1+d)}{4-d^2}r^F\right], \qquad q_2^{F*} = \frac{1}{1+d}\left[\frac{(A-c)}{2-d} - \frac{2(1+d)}{4-d^2}r^F\right]$$

Since $q_2^{N*} < q_1^{N*}$ for $0 < d < 1$, we have an interior solution if

$$q_2^{N*} = \frac{1}{1+d}\left[\frac{(A-c)}{2-d} - \frac{2(1+d)}{4-d^2}(r^F + s)\right] > 0$$

or

$$r^F + s < \frac{(2+d)}{2(1+d)}(A-c),$$

validating the discussion in footnote 26.

Let

$$\pi_1^{F*} = (p_1^{F*} - c_1)q_1^{F*} + r^F q_2^{F*}$$

be Qualcomm's profits when it collects only FRAND royalty $r^F$ and

$$\pi_1^{N*} = (p_1^{N*} - c_1)q_1^{N*} + (r^F + s)q_2^{N*}$$

be Qualcomm's profits when it imposes a non-discriminatory royalty surcharge $s$ on top of the FRAND royalty.

I show that

$$\pi_1^{N*} > \pi_1^{F*} \; for \; 0 < d < 1 \; if \; and \; only \; if \; 2r^F + s < \frac{8+d^3}{8+d^2}(A-c)$$

First define $S \equiv \frac{1}{4-d^2}s$. Then

$$\pi_1^{N*} = (p_1^{N*} - c_1)q_1^{N*} + (r^F + s)q_2^{N*}$$

$$= (p_1^{F*} - c_1 + 3dS)(q_1^{F*} - dS) + r^F q_2^{F*} - 2r^F S + s(q_2^{F*} - 2S)$$

$$= \pi_1^{F*} + 3dS(q_1^{F*} - dS) - (p_1^{F*} - c_1)dS - 2r^F S + (4-d^2)S(q_2^{F*} - 2S)$$

so that

$$\frac{\pi_1^{N*} - \pi_1^{F*}}{S} = 3d(q_1^{F*} - dS) - (p_1^{F*} - c_1)d - 2r^F + (4-d^2)(q_2^{F*} - 2S)$$

Substituting from firm 1's FOC, $p_1^{F*} - c_1 - dr^F = (1-d^2)q_1^{F*}$:

$$= d(2+d^2)q_1^{F*} + (4-d^2)q_2^{F*} - (2+d^2)r^F - [8+d^2]S$$

$$= (4 + 2d - d^2 + d^3)\frac{(A-c)}{(1+d)(2-d)} - \left[\frac{d^2(2+d^2)}{4-d^2} + (4+d^2)\right]r^F - [8+d^2]\frac{1}{4-d^2}s$$

$$= \frac{1}{(4-d^2)}[(8+d^3)(A-c) - (8+d^2)(2r^F + s)] > 0$$

if and only if $2r^F + s < \frac{8+d^3}{8+d^2}(A-c)$. ■

*Proof of Proposition* 5. The proof compares the comparative-statics formulas $\left.\frac{dp_i^{N*}}{ds}\right|_{s=0}$ and $\left.\frac{dp_i^{G*}}{dt}\right|_{t=0}$ derived from the surcharge and tax FOCs respectively. Let $det\Pi \equiv \pi_{11}^1\pi_{22}^2 - \pi_{12}^1\pi_{12}^2 > 0$ under (A5) and $\Delta \equiv \left.\frac{(q_1^1+q_1^2)\pi_{22}^2}{det\Pi}\right|_{s=0} > 0$ under (A3) and (A5).

I first show that $\left.\frac{dp_1^{G*}}{dt}\right|_{t=0} - \left.\frac{dp_1^{N*}}{ds}\right|_{s=0} = \Delta > 0$. I then show that $\left.\frac{dp_2^{G*}}{dt}\right|_{t=0} - \left.\frac{dp_2^{N*}}{ds}\right|_{s=0} = \Delta \times \left.\left(\frac{dp_2^G(p_1^G)}{dp_1^G}\right)\right|_{s=0} < \Delta$.

The two systems' second-derivative matrices are for the tax case,

$$\pi_{11}^1(p_1^G, p_2^G) = 2q_1^1 + (p_1^G - c_1 - t)q_{11}^1 + r^F q_{11}^2$$

$$\pi_{12}^1(p_1^G, p_2^G) = q_2^1 + (p_1^G - c_1 - t)q_{12}^1 + r^F q_{12}^2$$

$$\pi_{22}^2(p_1^G, p_2^G) = 2q_2^2 + (p_2^G - c_2 - r^F - t)q_{22}^2$$

$$\pi_{12}^2(p_1^G, p_2^G) = q_1^2 + (p_2^G - c_2 - r^F - t)q_{12}^2$$

and for the surcharge case,

$$\pi_{11}^1(p_1^N, p_2^N) = 2q_1^1 + (p_1^N - c_1)q_{11}^1 + (r^F + s)q_{11}^2$$

$$\pi_{12}^1(p_1^N, p_2^N) = q_2^1 + (p_1^N - c_1)q_{12}^1 + (r^F + s)q_{12}^2$$

$$\pi_{22}^2(p_1^N, p_2^N) = 2q_2^2 + (p_2^N - c_2 - r^F - s)q_{22}^2$$

$$\pi_{12}^2(p_1^N, p_2^N) = q_1^2 + (p_2^N - c_2 - r^F - s)q_{12}^2$$

Evaluating both sets of partials at $s = t = 0$, where $p_i^N = p_i^G = p_i^F$, yields

$$\pi_{ij}^i(p_1^N, p_2^N) = \pi_{ij}^i(p_1^G, p_2^G) = \pi_{ij}^i(p_1^F, p_2^F)$$

The four cross-partials are therefore the same in both systems at $s = t = 0$. From Cramer's rule applied to the tax FOCs:

$$\left.\frac{dp_1^{G*}}{dt}\right|_{t=0} = \left.\frac{\begin{vmatrix} q_1^1 & \pi_{12}^1 \\ q_2^2 & \pi_{22}^2 \end{vmatrix}}{det\Pi}\right|_{t=0}$$

Applying the same procedure to the surcharge FOCs, with the right-hand side $(-q_1^2, q_2^2)$ instead of $(q_1^1, q_2^2)$:

$$\left.\frac{dp_1^{N*}}{ds}\right|_{s=0} = \left.\frac{\begin{vmatrix} -q_1^2 & \pi_{12}^1 \\ q_2^2 & \pi_{22}^2 \end{vmatrix}}{det\Pi}\right|_{s=0} = \left.\frac{\begin{vmatrix} q_1^1 & \pi_{12}^1 \\ q_2^2 & \pi_{22}^2 \end{vmatrix}}{det\Pi}\right|_{s=0} + \left.\frac{\begin{vmatrix} -(q_1^1 + q_1^2) & \pi_{12}^1 \\ 0 & \pi_{22}^2 \end{vmatrix}}{det\Pi}\right|_{s=0}$$

$$= \left.\frac{dp_1^{G*}}{dt}\right|_{t=0} - \Delta$$

The argument for firm 2 follows the same template, with one twist: the numerator of $\left.\frac{dp_2^{N*}}{ds}\right|_{s=0}$ depends on $-q_1^2$ rather than $q_2^2$.

$$\left.\frac{dp_2^{G*}}{dt}\right|_{t=0} = \left.\frac{\begin{vmatrix}\pi_{11}^1 & q_1^1 \\ \pi_{12}^2 & q_2^2\end{vmatrix}}{det\Pi}\right|_{t=0}$$

and

$$\left.\frac{dp_2^{N*}}{ds}\right|_{s=0} = \left.\frac{\begin{vmatrix}\pi_{11}^1 & -q_1^2 \\ \pi_{12}^2 & q_2^2\end{vmatrix}}{det\Pi}\right|_{s=0} = \left.\frac{dp_2^{G*}}{dt}\right|_{t=0} + \left.\frac{(q_1^1 + q_1^2)\pi_{12}^2}{det\Pi}\right|_{s=0}$$

$$= \left.\frac{dp_2^{G*}}{dt}\right|_{t=0} - \Delta \times \left.\left(\frac{dp_2^G(p_1^G)}{dp_1^G}\right)\right|_{s=t=0}$$

where the last equality uses firm 2's best-response slope $\frac{dp_2^G(p_1^G)}{dp_1^G} = -\frac{\pi_{12}^2}{\pi_{22}^2}$, which is positive and less than one under (A5). ■

*Proof of Proposition* 6. Under Qualcomm's Implicit Assumption, $\left.\frac{dp_1^{G*}}{dt}\right|_{t=0} = \left.\frac{dp_2^{G*}}{dt}\right|_{t=0}$ so that the inequality in Proposition 5 becomes $\left.\frac{dp_1^{N*}}{ds}\right|_{s=0} < \left.\frac{dp_2^{N*}}{ds}\right|_{s=0}$. The lower bound $0 < \left(\frac{dp_1^{N*}}{ds}\right)_{s=0}$ is the surcharge pass-through result established in Proposition 2. ■

*Proof of Proposition* 7.

$$\left.\frac{dq_2^{N*}}{ds}\right|_{s=0} = q_1^2 \times \left.\frac{dp_1^{N*}}{ds}\right|_{s=0} + q_2^2 \times \left.\frac{dp_2^{N*}}{ds}\right|_{s=0} < (q_1^2 + q_2^2) \times \left.\frac{dp_2^{N*}}{ds}\right|_{s=0} < 0,$$

where the first inequality follows from (A2) and Propositions 2 and 6 and the second inequality from (A4) and Proposition 2. ■

*Proof of Proposition* 8.

$$\left.\frac{dq_1^{N*}}{ds}\right|_{s=0} - \left.\frac{dq_2^{N*}}{ds}\right|_{s=0} = \left.\left(q_1^1\frac{dp_1^{N*}}{ds} + q_2^1\frac{dp_2^{N*}}{ds}\right)\right|_{s=0} - \left.\left(q_1^2\frac{dp_1^{N*}}{ds} + q_2^2\frac{dp_2^{N*}}{ds}\right)\right|_{s=0}$$

$$= (q_1^1 - q_1^2)\left.\frac{dp_1^{N*}}{ds}\right|_{s=0} - (q_2^2 - q_2^1)\left.\frac{dp_2^{N*}}{ds}\right|_{s=0}$$

$$= (q_1^1 - q_1^2)|_{t=0} \times \left[\left.\frac{dp_1^{G*}}{dt}\right|_{t=0} - \Delta\right]$$

$$-(q_2^2 - q_2^1)|_{t=0} \times \left\{ \left.\frac{dp_2^{G*}}{dt}\right|_{t=0} - \Delta \times \left.\frac{dp_2^G(p_1^G)}{dp_1^G}\right|_{s=t=0} \right\}$$

where the last equality follows from the proof of Proposition 5 and the fact that $q_i^j|_{t=0} = q_i^j|_{s=0}, i,j = 1,2$. Qualcomm's Implicit Assumption only guarantees that $\left.\frac{dp_1^{G*}}{dt}\right|_{t=0} = \left.\frac{dp_2^{G*}}{dt}\right|_{t=0}$. Without further structure, the sign of this expression is ambiguous. Under (A4′), however, the two coefficients $(q_1^1 - q_1^2)$ and $(q_2^2 - q_2^1)$ coincide, and the two bracketed terms differ only by the BR-slope factor inside the second one. Therefore

$$\left.\frac{dq_1^{N*}}{ds}\right|_{s=0} - \left.\frac{dq_2^{N*}}{ds}\right|_{s=0} = -(q_1^1 - q_1^2)|_{t=0} \times \Delta \left\{ 1 - \left.\frac{dp_2^G(p_1^G)}{dp_1^G}\right|_{s=t=0} \right\} > 0 \qquad \blacksquare$$

## Appendix B: Ad-Valorem Surcharges on Handset Prices

**[To be provided]**